\documentclass[reprint,amsmath,amssymb,aps,pra,longbibliography]{revtex4-2}

\usepackage{graphicx}
\usepackage{dcolumn}
\usepackage{bm}
\usepackage{xcolor}
\usepackage{mathptmx}
\usepackage{comment}
\usepackage{array}[=2019-01-01]

\begin{document}

\title{Limits on the free-space group velocity of optical wave packets incorporating angular dispersion. Part~II, non-differentiable angular dispersion: tutorial}

\author{Layton A. Hall$^{1,2}$}
\author{Ayman F. Abouraddy$^1$}
\email{raddy@creol.ucf.edu}
\affiliation{$^1$CREOL, The College of Optics \& Photonics, University of Central Florida, Orlando, FL 32816, USA}
\affiliation{$^2$Los Alamos National Laboratory, Los Alamos, NM 87545, USA}

\begin{abstract}
In Part~I of this tutorial, we showed that introducing angular dispersion (AD) into a collimated optical pulse allows for tuning the wave packet group velocity in free space. In principle, the group velocity can take on arbitrary values, whether superluminal, subluminal, or even negative. However, any significant deviation of the group velocity from $c$ (the speed of light in vacuum) necessitates propagation at large angles with respect to the optical axis in the non-paraxial regime. In Part~II of this tutorial, we describe the recent discovery of a new form of AD we refer to as `non-differentiable AD', which circumvents the limits imposed by conventional (differentiable) AD. Non-differentiable AD does \textit{not} refer to an AD profile that is discontinuous, contains a kink, or features a singularity. Rather, a non-differentiable AD profile is continuous, smooth, and is differentiable everywhere except for one wavelength at which the derivative is not defined. We show that salutary features follow from this non-differentiability with respect to tuning the group velocity of a pulsed beam in free space; namely, significant deviations in the group velocity away from $c$ are achievable in the paraxial regime while remaining free of group-velocity dispersion and all higher-order dispersive effects. This tutorial presents an outline of these recent results for tuning the group velocity of a wave packet via non-differentiable AD, connecting them with the corresponding results associated with conventional AD -- as outlined in Part~I. Moreover, we show that non-differentiable AD undergirds the unique characteristics of propagation-invariant space-time wave packets, thus unifying a large swathe of results in a single conceptual framework.
\end{abstract}

\maketitle

\section{Introduction}

We reviewed in Part~I of this tutorial \cite{Hall26JOSAA1} the concept of angular dispersion (AD) and how introducing AD into a pulsed beam (or wave packet) enables tuning its group velocity $\widetilde{v}$ in free space away from $c$ (the speed of light in vacuum) when measured along the optical axis (taken to be the $z$-axis). Although the group velocity of an AD-endowed wave packet can take on -- in principle -- arbitrary values (whether subluminal, superluminal, or even negative), several factors nevertheless prevent such effects from being practically realized:
\begin{enumerate}
\item A significant deviation of $\widetilde{v}$ from $c$ requires a large propagation angle (i.e., operating in the non-paraxial regime), in which case the new group velocity can be observed only over a short distance.
\item Operating at smaller propagation angles requires independent control over both the propagation angle and the linear AD coefficient, in addition to access to large values of this coefficient. Most optical configurations that introduce AD do not satisfy these requirements.
\item An AD-endowed wave packet experiences group velocity dispersion (GVD) in free space, thereby setting an upper limit on the distance over which the group velocity $\widetilde{v}$ can be monitored.
\end{enumerate}

Our goal in Part~II of this tutorial is to provide an overview of the new concept of `non-differentiable AD' that helps circumvent the above-listed limits on the tunability of the group velocity of a pulsed optical beam in free space when relying on conventional AD. To this end, we obtain limits on the achievable deviation in $\widetilde{v}$ from $c$ in free space for AD-endowed wave packets (whether differentiable or non-differentiable AD) subject to a given available numerical aperture (NA) and a selected pulse bandwidth. We find that non-differentiable AD allows for dramatic tunability of $\widetilde{v}$ while remaining in the paraxial regime. The origin of this novel effect is found mathematically in the role played by the bandwidth in determining the NA. In the case of conventional (differentiable) AD, the bandwidth enters as an \textit{additive} term, and neglecting it does not change the numerical aperture significantly. By contrast, in the case of non-differentiable AD, the bandwidth enters as an overall \textit{multiplicative} term, so that reducing the bandwidth allows for a dramatic reduction in the required numerical aperture. Additionally, we determine the maximum group delay that an AD-endowed pulsed beam displays relative to a reference plane-wave pulse, which is limited by \textit{diffractive} spatial spreading and \textit{dispersive} temporal spreading that inevitably accompany conventional AD. We show that non-differentiable AD provides additional advantages in this regard, namely, eliminating diffraction for extended distances \textit{and} the AD-induced GVD. The formulation presented here therefore explains why previous experiments (prior to $\sim2017$) revealed only minute deviations in $\widetilde{v}$ from $c$, and why recent experiments have vastly superseded them.

Part~II of this tutorial is structured as follows. After a brief review of the salient results regarding the group velocity of wave packets endowed with conventional AD from Part~I \cite{Hall26JOSAA1}, we describe the concept of non-differentiable AD and derive it from the requirement of tuning the group velocity $\widetilde{v}$ away from $c$ along the optical axis in free space. We then connect the concept of non-differentiable AD with recently developed, propagation-invariant space-time wave packets (STWPs). Next, we obtain bounds for the minimum NA needed to produce a wave packet of given bandwidth and target group velocity. This bound allows us to elucidate why non-differentiable AD enables significant changes in $\widetilde{v}$ in the paraxial regime when compared to conventional (differentiable) AD. Finally, we discuss practical bounds on the observability of group-velocity changes in free space, and provide a comparison between the concepts associated with AD and those associated with chromatic dispersion.

\section{Tuning the group velocity by introducing angular dispersion}

In Part~I of this tutorial \cite{Hall26JOSAA1}, we obtained a general expression for the group velocity $\widetilde{v}$ of a wave packet endowed with AD. This expression presumes only that the propagation angle $\varphi(\omega)$ for the frequency $\omega$ with respect to a fixed direction (here the $z$-axis) can be expanded perturbatively around a fixed frequency $\omega_{\mathrm{o}}$:
\begin{equation}\label{eq:ADexpansion}
\varphi(\omega)=\varphi(\omega_{\mathrm{o}}+\Omega)\approx\varphi_{\mathrm{o}}+\varphi_{\mathrm{o}}^{(1)}\Omega+\frac{1}{2}\varphi_{\mathrm{o}}^{(2)}\Omega^{2}+\cdots,
\end{equation}
where $\Omega=\omega-\omega_{\mathrm{o}}$,  $\varphi_{\mathrm{o}}=\varphi(\omega_{\mathrm{o}})$ is the propagation angle for $\omega_{\mathrm{o}}$, $\varphi_{\mathrm{o}}^{(1)}=\tfrac{d\varphi}{d\omega}\bigr|_{\omega_{\mathrm{o}}}$, and $\varphi_{\mathrm{o}}^{(2)}=\tfrac{d^{2}\varphi}{d\omega^{2}}\big|_{\omega_{\mathrm{o}}}$ \cite{Porras03PRE2,Hall25APLP}. We define on-axis propagation to be associated with wave packets in which $\varphi_{\mathrm{o}}=0$ and off-axis propagation with $\varphi_{\mathrm{o}}\neq0$. Within this model, the group velocity $\widetilde{v}$ along the $z$-axis in free space is given by:
\begin{equation}\label{eq:GroupVelocityGeneral}
\widetilde{v}=\frac{c}{\cos\varphi_{\mathrm{o}}-\omega_{\mathrm{o}}\varphi_{\mathrm{o}}^{(1)}\sin\varphi_{\mathrm{o}}}.
\end{equation}
Contrary to the intuitive notion that spatially structuring a collimated pulse necessarily reduces the group velocity \cite{Sambles15Science,Giovannini15Science,Bouchard16Optica}, this expression indicates that $\widetilde{v}$ can take on arbitrary values in free space, whether subluminal, superluminal or negative. However, several desiderata are needed to exercise full control over $\widetilde{v}$:
\begin{enumerate}
\item off-axis propagation $\varphi_{\mathrm{o}}\neq0$ is needed;
\item $\varphi_{\mathrm{o}}$ and $\varphi_{\mathrm{o}}^{(1)}$ need to be independently controllable; and
\item arbitrary values of $\varphi_{\mathrm{o}}^{(1)}$ must be accessible.
\end{enumerate}
These requirements are difficult to satisfy in traditional optical settings \cite{Hall25APLP}. Consequently, only minute changes in $\widetilde{v}$ with respect to $c$ have been realized to date in free space by introducing AD. Furthermore, an AD-endowed wave packet is necessarily dispersive; i.e., it experiences GVD in free space and thus disperses temporally with propagation distance \cite{Martinez84JOSAA}. Eliminating GVD is only possible for an off-axis wave packet when independent control over $\varphi_{\mathrm{o}}$, $\varphi_{\mathrm{o}}^{(1)}$, \textit{and} $\varphi_{\mathrm{o}}^{(2)}$ is available \cite{Porras03PRE2,Zapata06OL,Hall25APLP,Hall26JOSAA1}. It is impossible to eliminate AD-induced GVD for an on-axis wave packet endowed with conventional AD.

\section{Non-differentiable angular dispersion}

\subsection{What is non-differentiable AD?}

The crucial assumption underpinning Eq.~\ref{eq:GroupVelocityGeneral} is the \textit{differentiability} of $\varphi(\omega)$ at $\omega=\omega_{\mathrm{o}}$; i.e., $\varphi_{\mathrm{o}}^{(1)}=\tfrac{d\varphi}{d\omega}\big|_{\omega_{\mathrm{o}}}$ is well-defined. Consider the counter scenario where $\varphi(\omega)$ is \textit{not} differentiable at $\omega_{\mathrm{o}}$. This does not require that $\varphi(\omega)$ have a discontinuity, a kink, or a singularity at $\omega_{\mathrm{o}}$. Indeed, $\varphi(\omega)$ has only a finite domain $[0,2\pi]$, so that singularities do not exist in $\varphi(\omega)$, although discontinuities or kinks may arise. Rather, consider the spectral AD profile:
\begin{equation}\label{eq:ExampleOfNonDiffAD}
\varphi(\omega)\propto\sqrt{\omega-\omega_{\mathrm{o}}},
\end{equation}
with $\omega>\omega_{\mathrm{o}}$ [Fig.~\ref{fig:NonDiffAD}(a)]. This profile is continuous, finite everywhere, and differentiable everywhere \textit{except} at $\omega=\omega_{\mathrm{o}}$. Because $\sqrt{x}$ is not differentiable at $x=0$, $\varphi_{\mathrm{o}}^{(1)}$ is not defined for the AD profile in Eq.~\ref{eq:ExampleOfNonDiffAD}. We refer to this form of $\varphi(\omega)$ as `non-differentiable AD', and we call $\omega_{\mathrm{o}}$ the non-differentiable frequency \cite{Hall25APLP}. The consequences of introducing non-differentiable AD for the tunability of the group velocity $\widetilde{v}$ are profound \cite{Hall21OL,Hall22OEConsequences,Hall22JOSAA,Hall25APLP}, and the well-established findings regarding the group velocity that rely on Eq.~\ref{eq:ADexpansion} are overturned in presence of non-differentiable AD \cite{Hall21OLNormalGVD,Hall23LPR}, as we proceed to demonstrate.

\subsection{Non-differentiable AD for on-axis propagation}

Let us approach the problem of an \textit{on-axis} field, whereupon $\omega\rightarrow\omega_{\mathrm{o}}$ is associated with $\varphi\rightarrow\varphi_{\mathrm{o}}=0$. We start from the general expression for the group velocity of an AD-endowed wave packet in Eq.~\ref{eq:GroupVelocityGeneral} and then take the small-angle approximation ($\cos\varphi\approx1$ and $\sin\varphi\approx\varphi$) that is appropriate in the vicinity of $\omega=\omega_{\mathrm{o}}$ and $\varphi=\varphi_{\mathrm{o}}=0$, whereupon:
\begin{equation}
\widetilde{v}\approx\frac{c}{1-\left(\omega\varphi\tfrac{d\varphi}{d\omega}\right)\bigr|_{\omega_{\mathrm{o}}}}.
\end{equation}
Our aim now is to find the condition for $\widetilde{v}\neq c$. This requires of course that $\left(\omega\varphi\tfrac{d\varphi}{d\omega}\right)\bigr|_{\omega_{\mathrm{o}}}\neq0$, which seems impossible because $\varphi\rightarrow0$ when $\omega\rightarrow\omega_{\mathrm{o}}$. However, this traditional conclusion assumes that $\tfrac{d\varphi}{d\omega}\bigr|_{\omega_{\mathrm{o}}}$ is finite.

\begin{figure}[t!]
\centering
\includegraphics[width=8.8cm]{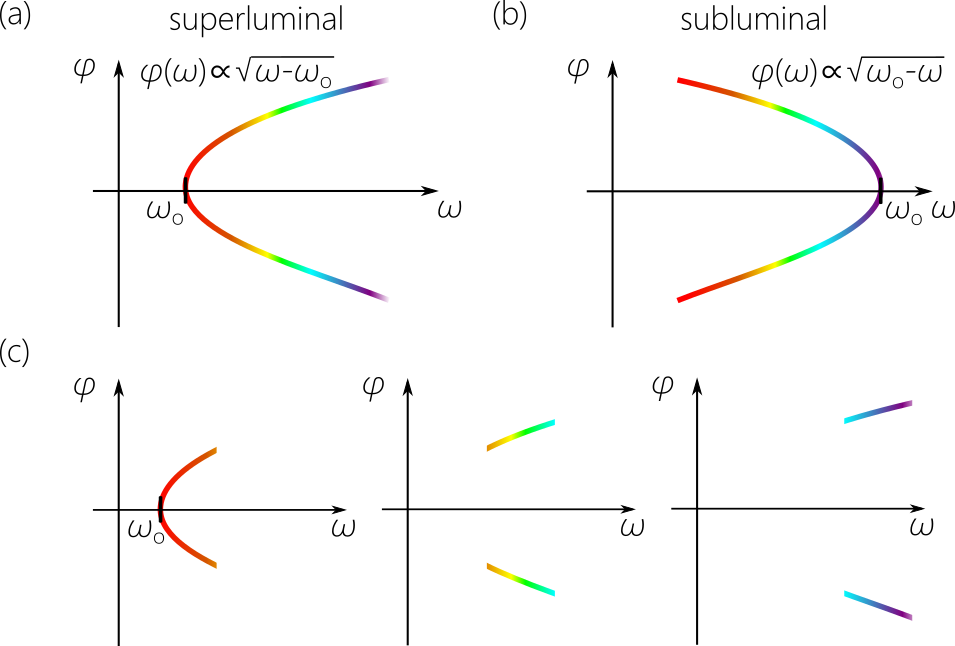}
\caption{Spectral AD profiles associated with non-differentiable AD corresponding to (a) superluminal group velocity where $\varphi(\omega)\propto\sqrt{\omega-\omega_{\mathrm{o}}}$, and (b) subluminal group velocity where $\varphi(\omega)\propto\sqrt{\omega_{\mathrm{o}}-\omega}$. (c) On-axis (left) and off-axis (middle and right) versions of the AD profile in (a). In contradistinction to differentiable AD, in which case on-axis and off-axis wave packets have different group velocities (even in the presence of the same AD) \cite{Hall26JOSAA1}, here all three wave packets have the same group velocity \cite{Yessenov19PRA,Hall22JOSAA}.}%\vspace{-5mm}
\label{fig:NonDiffAD}
\end{figure}

Let us now set the quantity $\varphi\tfrac{d\varphi}{d\omega}$ to be a frequency-independent constant, $\varphi\tfrac{d\varphi}{d\omega}=\tfrac{\eta}{\omega_{\mathrm{o}}}$, where $\eta>0$ is a dimensionless constant, which allows us to achieve the goal of changing $\widetilde{v}$ for this on-axis wave packet. By integration we obtain:
\begin{equation}\label{eq:SuperluminalNonDiffAD}
\int_{\varphi_{\mathrm{o}}=0}^{\varphi}\!\!\varphi\;d\varphi=\frac{\eta}{\omega_{\mathrm{o}}}\int_{\omega_{\mathrm{o}}}^{\omega}\!\!d\omega\;\;\longrightarrow\;\;\varphi(\omega)=\sqrt{2\eta\frac{\omega-\omega_{\mathrm{o}}}{\omega_{\mathrm{o}}}},
\end{equation}
with the restriction $\omega>\omega_{\mathrm{o}}$, which recovers $\varphi(\omega)$ in Eq.~\ref{eq:ExampleOfNonDiffAD} and Fig.~\ref{fig:NonDiffAD}(a). Note that $\varphi\rightarrow0$ when $\omega\rightarrow\omega_{\mathrm{o}}$ is associated with $\tfrac{d\varphi}{d\omega}\rightarrow\infty$, and $\omega\varphi\tfrac{d\varphi}{d\omega}\big|_{\omega_{\mathrm{o}}}=\eta$. Consequently, $\widetilde{v}=\tfrac{c}{1-\eta}$ and the group index is $\widetilde{n}=1-\eta$. This field configuration results in $\widetilde{v}>c$ (superluminal) when $0<\eta<1$, and $\widetilde{v}<0$ (negative-$\widetilde{v}$) when $\eta>1$. From this, we have circumscribed a fundamental limitation of conventional AD: we now have $\widetilde{v}\neq c$ \textit{for on-axis propagation} ($\varphi_{\mathrm{o}}=0$). Alternatively, if we set $\varphi\tfrac{d\varphi}{d\omega}=-\tfrac{\eta}{\omega_{\mathrm{o}}}$, with $\eta>0$, then:
\begin{equation}\label{eq:SubluminalNonDiffAD}
\varphi(\omega)=\sqrt{2\eta\frac{\omega_{\mathrm{o}}-\omega}{\omega_{\mathrm{o}}}},
\end{equation}
where $\omega<\omega_{\mathrm{o}}$ [Fig.~\ref{fig:NonDiffAD}(b)]. We again have non-differentiable AD: $\tfrac{d\varphi}{d\omega}\bigr|_{\omega_{\mathrm{o}}}$ is not defined, $\omega\varphi\tfrac{d\varphi}{d\omega}=-\eta$ independently of $\omega$, and $\widetilde{v}=\tfrac{c}{1+\eta}<c$ is subluminal ($\widetilde{n}=1+\eta$) for all values of $\eta$.

The AD profile in Fig.~\ref{fig:NonDiffAD}(a) is symmetrized with respect to $\varphi_{\mathrm{o}}$, so that the associated wave packet has an on-axis intensity peak. The group velocity $\widetilde{v}$ is simply the speed of this intensity peak along the optical axis. Of course, one may object to the characterization of this wave packet as `on-axis' and instead consider it an `off-axis' wave packet with $\varphi_{\mathrm{o}}$ located somewhere between $\varphi=0$ and $\varphi_{\mathrm{max}}=\sqrt{2\eta\tfrac{\Delta\omega}{\omega_{\mathrm{o}}}}$, where $\Delta\omega$ is the bandwidth.  Because of the rapid variation in $\varphi$ with $\omega$, it is not clear how to choose the new $\varphi_{\mathrm{o}}$ (by contrast to conventional AD that is typically dominated by the linear AD term, thus making identifying $\varphi_{\mathrm{o}}$ more straightforward \cite{Hall26JOSAA1}). More importantly, as we show below, the group velocity can be tuned dramatically away from $c$ for vanishingly small $\varphi_{\mathrm{o}}$ in contradistinction to conventional (differentiable) AD. We thus maintain the `on-axis' characterization for the AD profile in Fig.~\ref{fig:NonDiffAD}(a,b) to distinguish it from cases where the spectrum is shifted along this parabolic curve by eliminating the spectrum in the vicinity of $\varphi=0$ [Fig.~\ref{fig:NonDiffAD}(c)]. This reveals a unique feature of non-differentiable AD: \textit{the realized group velocity does not change while moving a fixed bandwidth along the parabolic curve moving from an on-axis to an off-axis configuration}. This is a consequence of the product $\varphi\tfrac{d\varphi}{d\omega}$ being a frequency-independent constant, a condition not satisfied by any other form of AD [Fig.~\ref{fig:NonDiffAD}(c)].

The rapid variation in $\varphi$ with $\omega$ in the vicinity of the non-differentiable frequency $\omega_{\mathrm{o}}$ makes defining the phase velocity $v_{\mathrm{ph}}$ difficult. We have previously taken the phase velocity of a wave packet endowed with non-differentiable AD to be $v_{\mathrm{ph}}=\tfrac{\omega}{k_{z}}\big|_{\omega_{\mathrm{o}}}=c$ for consistency with wave packets endowed with conventional AD \cite{Kondakci17NP,Kondakci19NC}. Nevertheless, more work is needed to determine a definition of phase velocity in presence of non-differentiable AD, which is important in applications involving phase matching in nonlinear optics.

\subsection{Salutary features of non-differentiable AD}

It is thus clear that non-differentiable AD violates the rule previously demonstrated for differentiable AD whereby $\widetilde{v}=c$ for on-axis wave packets in free space \cite{Hall26JOSAA1} (Eq.~\ref{eq:GroupVelocityGeneral}). By contrast, \textit{non-differentiable AD} enables tuning $\widetilde{v}$ arbitrarily on-axis, which is accompanied by two other attractive features (as detailed below):
\begin{enumerate}
\item Dramatic deviations in $\widetilde{v}$ from~$c$ in free space are realizable in the paraxial regime.
\item The wave packet is free of AD-induced GVD and all higher-order dispersive effects. 
\end{enumerate}

The first feature can be readily recognized by an example. Consider a pulse of bandwidth $\Delta\lambda=1$~nm at $\lambda_{\mathrm{o}}=800$~nm (pulse width $\sim1$~ps). A superluminal group velocity $\widetilde{v}=2c$ ($\eta=\tfrac{1}{2}$ in Eq.~\ref{eq:SuperluminalNonDiffAD}) -- which is a dramatic departure from $c$ -- is realized with non-differentiable AD accompanying a maximum angle $\varphi\approx2^{\circ}$, compared with $\varphi_{\mathrm{o}}=60^{\circ}$ for an X-wave \cite{Saari97PRL,Yessenov19PRA,Hall26JOSAA1}. Similarly in the subluminal regime, a group velocity $\widetilde{v}=0.5c$ ($\eta=1$ in Eq.~\ref{eq:SubluminalNonDiffAD}) requires a maximum angle of $\varphi\approx2^{\circ}$, compared with a pulsed Bessel beam that requires $\varphi_{\mathrm{o}}=60^{\circ}$ \cite{Giovannini15Science,Hall26JOSAA1}. These examples indicate clearly the surprising improvement brought about by non-differentiable AD over conventional AD (e.g., X-waves and pulsed Bessel beams) with respect to the required NA for a target group velocity at a fixed bandwidth. We discuss this feature further in Section~\ref{sec:NA} below, and we address the absence of AD-induced GVD in the next Section.

%Larger angle $k_{z}(\omega)=k_{\mathrm{o}}+(\omega-\omega_{\mathrm{o}})/\widetilde{v}$

%Synthesis. which can be readily prepared using the universal AD synthesizer in Refs.~\cite{Hall24JOSAA,Romer25JOpt}. 1D+2D \cite{Yessenov25Meron,Yessenov22NC}

%Propagation invariance

\section{Space-time wave packets}

\subsection{Connection between STWPs and non-differentiable AD}

The examples of non-differentiable AD in Eq.~\ref{eq:SuperluminalNonDiffAD} (superluminal, $\omega>\omega_{\mathrm{o}}$) and in Eq.~\ref{eq:SubluminalNonDiffAD} (subluminal, $\omega<\omega_{\mathrm{o}}$) are valid only for small angles. However, these profiles are approximations of a more general constraint for a wave packet whose axial wave number $k_{z}$ is linearly proportional to the frequency $\omega$ \cite{Kondakci17NP,Yessenov19PRA,Yessenov22AOP}:
\begin{equation}\label{eq:STWPconstraint}
\omega-\omega_{\mathrm{o}}=(k_{z}-k_{\mathrm{o}})\widetilde{v}.
\end{equation}
This constraint corresponds to a hyperplane in spectral space $(k_{x},k_{y},k_{z},\tfrac{\omega}{c})$ making an angle $\theta$ with the $k_{z}$-axis, whereupon the wave packet group velocity is $\widetilde{v}=c\tan\theta$. Optical fields characterized by such a constraint are known as space-time wave packets (STWPs), the unique class of propagation-invariant pulsed beams that maintains a constant group velocity \cite{Almeida25OL}. STWPs have demonstrated a host of other useful properties, including dispersion cancellation \cite{Hall23LPR,Hall23NatPhys,Hall24ACSP,Hall25PRA}, self-healing \cite{Kondakci18OL}, a multiplicity of space-time Talbot effects \cite{Yessenov20PRL,Hall21APLSTTalbot,Hall21OLTalbot}, and anomalous refraction \cite{Bhaduri20NP,Motz21OL,Yessenov21JOSAA3}, which have found applications in plasmonics \cite{Schepler20ACSPhot,Ichiji23PRA,Ichiji23ACSP,Ichiji24JOSAA,Ichiji25NC}, light-sheet microscopy \cite{Zhang25LSM}, imaging \cite{Diouf22SA}, line-of-sight target avoidance \cite{Hall26SubRayleigh,Hall26SelectiveAvoidance}, and interferometry \cite{Diouf23Optica}.

\begin{figure}[t!]
\centering
\includegraphics[width=8.8cm]{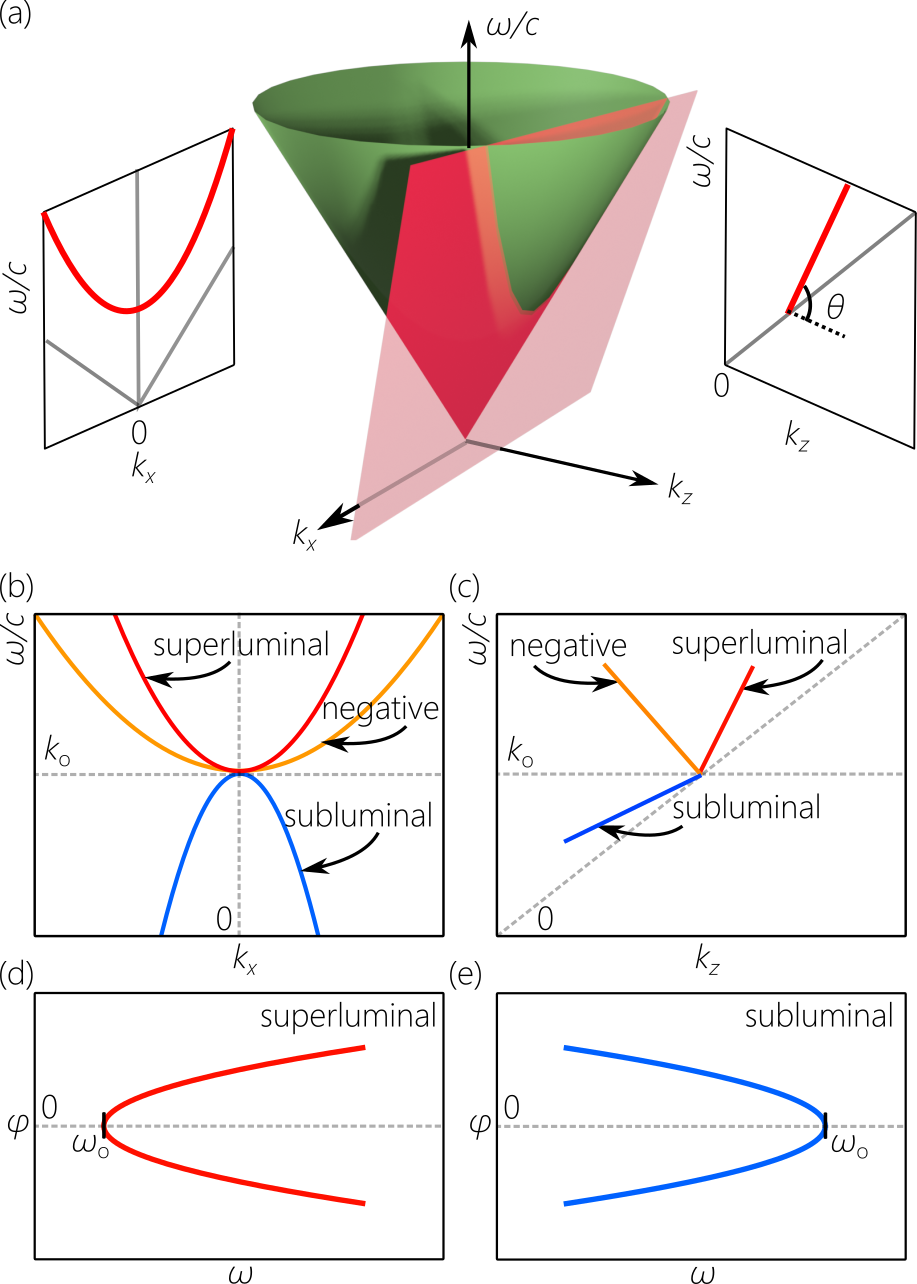}
\caption{(a) Depiction of the spectral domain for an STWP on the surface of the free-space light-cone in $(k_{x},k_{z},\tfrac{\omega}{c})$-space. This domain is the conic section at the intersection of the light-cone with a tilted spectral plane, whose projection onto the $(k_{z},\tfrac{\omega}{c})$-plane is a straight line. The slope of this projection determines the group velocity $\widetilde{v}$ of the STWP. (b) The spectral projection on the $(k_{x},\tfrac{\omega}{c})$-plane is a conic section. In the vicinity of $k_{x}=0$, any such conic section can be approximated by a parabola. In the small-angle approximation, this parabola matches the non-differentiable AD profile in Fig.~\ref{fig:NonDiffAD}(a,b). We plot three projections associated with subluminal, superluminal, and negative-$\widetilde{v}$ STWPs. (c) Projections onto the $(k_{z},\tfrac{\omega}{c})$-plane associated with those in (b). The tilted dashed line in the free-space light-line $k_{z}=\omega/c$. (d) The AD profile associated with a superluminal STWP extracted from the spectral projection onto the $(k_{x},\tfrac{\omega}{c})$-plane in (b); see Fig.~\ref{fig:NonDiffAD}(a). (e) Same as (d) for a subluminal STWP; see Fig.~\ref{fig:NonDiffAD}(b).}%\vspace{-5mm}
\label{fig:STWP}
\end{figure}

In free space, the dispersion relationship $k_{x}^{2}+k_{y}^{2}+k_{z}^{2}=(\tfrac{\omega}{c})^{2}$, or the light hyper-cone, defines all possible propagating monochromatic plane waves. Confining ourselves either to fields that are uniform along~$y$, so that $k_{y}=0$ and $k_{x}^{2}+k_{z}^{2}=(\tfrac{\omega}{c})^{2}$, or fields that are circularly symmetric, so that $k_{r}^{2}=k_{x}^{2}+k_{y}^{2}$ and $k_{r}^{2}+k_{z}^{2}=(\tfrac{\omega}{c})^{2}$, the spectral support associated with the constraint in Eq.~\ref{eq:STWPconstraint} is a conic section at the intersection of the light-cone with this spectral plane [Fig.~\ref{fig:STWP}(a)]. This conic section is an ellipse when $0<\theta<45^{\circ}$ or $135^{\circ}<\theta<180^{\circ}$, a hyperbola when $45^{\circ}<\theta<135^{\circ}$, a circle when $\theta=0^{\circ}$, a straight line when $\theta=45^{\circ}$, and a parabola when $\theta=135^{\circ}$ \cite{Yessenov19PRA}; see Fig.~\ref{fig:STWP}(b,c).

\subsection{AD-endowed STWPs are propagation invariant}

It is well-established that AD-endowed STWPs are propagation-invariant \cite{Kondakci16OE,Parker16OE,Kondakci17NP,Kondakci18OE,Yessenov22AOP}. Indeed, the field for an STWP can be written as $E(x,z;t)=e^{ik_{\mathrm{o}}(z-ct)}\psi(x,z;t)$, with a slowly varying spatiotemporal envelope $\psi(x,z;t)$ given by:
\begin{equation}\label{eq:STWPenvelope}
\psi(x,z;t)=\int\!d\Omega\widetilde{\psi}(\Omega)e^{i\{k_{x}x-\Omega(t-z/\widetilde{v})\}}=\psi(x,0;t-z/\widetilde{v}),
\end{equation}
which represents a wave packet traveling rigidly in free space without diffraction or dispersion at a group velocity $\widetilde{v}=c\tan\theta$.

In the paraxial regime $k_{z}(\omega)\approx\tfrac{\omega}{c}-\tfrac{k_{x}^{2}}{2k_{\mathrm{o}}}$, these conic sections can be approximated by a parabola in the vicinity of $k_{x}=0$:
\begin{equation}\label{eq:Parabola}
\frac{\omega-\omega_{\mathrm{o}}}{\omega_{\mathrm{o}}}\approx\frac{k_{x}^{2}}{2k_{\mathrm{o}}^{2}(1-\cot\theta)}.
\end{equation}
Note that $\omega>\omega_{\mathrm{o}}$ when $\theta>45^{\circ}$ (superluminal and negative-$\widetilde{v}$) and $\omega<\omega_{\mathrm{o}}$ when $\theta<45^{\circ}$ (subluminal). Writing $k_{x}(\omega)=\tfrac{\omega}{c}\sin\varphi(\omega)\approx\tfrac{\omega}{c}\varphi(\omega)$ yields $\varphi(\omega)\approx\sqrt{2\eta\tfrac{\omega-\omega_{\mathrm{o}}}{\omega_{\mathrm{o}}}}$, with $\eta=1-\cot\theta$ when $\theta>45^{\circ}$; or $\varphi(\omega)\approx\sqrt{2\eta\tfrac{\omega_{\mathrm{o}}-\omega}{\omega_{\mathrm{o}}}}$, with $\eta=\cot\theta-1$ when $\theta<45^{\circ}$. We can thus see that the forms of non-differentiable AD in Fig.~\ref{fig:NonDiffAD} correspond to small-angle approximations of STWPs.

Using pulses of width $\approx4.2$~ps at $\lambda_{\mathrm{o}}=800$~nm (bandwidth of $\Delta\lambda\approx0.3$~nm), the group velocity of the synthesized STWPs in Ref.~\cite{Kondakci19NC} were tuned from $30c$ in the superluminal regime, extending down to $0.5c$ in the subluminal regime, and to $-4c$ in the negative-$\widetilde{v}$ regime. Because $k_{x}$ and $\omega$ are tied to each other in Eq.~\ref{eq:Parabola} along with the group velocity $\widetilde{v}=c\tan\theta$, tuning $\widetilde{v}$ at fixed bandwidth $\Delta\omega$ therefore entails a concomitant change in the spatial bandwidth $\Delta k_{x}$ and thus also the transverse beam width $\Delta x\sim\tfrac{1}{\Delta k_{x}}$. For example, $\Delta x\approx17$~$\mu$m when $\widetilde{v}\approx1.34c$, and the beam width increases ($\Delta k_{x}$ or $\Delta\varphi$ decreases) as $\theta$ deviates away from $45^{\circ}$ ($\widetilde{v}$ deviates from $c$) for fixed bandwidth. The experimental synthesis of STWPs has been elaborated recently in detail, both for AD introduced along one transverse dimension \cite{Yessenov22AOP,Romer25JOpt} and for azimuthally symmetric fields \cite{Yessenov22NC,Yessenov22OL,Yessenov25JOSAA}.

\subsection{AD-endowed STWPs are GVD-free}

The spectral constraint characteristic of STWPs can be re-written as $k_{z}=k_{\mathrm{o}}+\tfrac{\omega-\omega_{\mathrm{o}}}{\widetilde{v}}$ (Eq.~\ref{eq:STWPconstraint}), whereupon $k_{z}$ is linear in $\omega$. Therefore $\tfrac{dk_{z}}{d\omega}=\tfrac{1}{\widetilde{v}}$ and $\tfrac{d^{2}k_{z}}{d\omega^{2}}=0$ at all frequencies; that is, STWPs are free from AD-induced GVD and all higher-order dispersive effects. The only constraint on the distance over which $\widetilde{v}$ is maintained arises from diffractive effects as we discuss below. That is, although STWPs are pulsed beams that are endowed with AD, they are nevertheless GVD-free in free space as a consequence of their non-differentiable AD profile. This is also clear from Eq.~\ref{eq:STWPenvelope} where the STWP envelope traveling rigidly in free space, and thus does not experience dispersive temporal spreading.

\section{Numerical aperture for a wave packet having a prescribed group velocity and bandwidth}\label{sec:NA}

\subsection{Numerical aperture for differentiable AD}

\subsubsection{Formulation of the problem}

As shown in Part~I of this tutorial \cite{Hall26JOSAA1} for the examples of the X-wave and the pulsed Bessel beam, a large deviation in $\widetilde{v}$ from $c$ via \textit{differentiable} AD typically requires a large propagation angle or, equivalently, a large NA. We aim here to establish more generally the minimum angle $\varphi_{\mathrm{min}}^{(\mathrm{D})}$ required to realize a prescribed group velocity $\widetilde{v}$ in free space for a given temporal bandwidth $\Delta\omega$ at a temporal frequency $\omega_{\mathrm{o}}$, so that $\mathrm{NA}=\sin\varphi_{\mathrm{min}}^{(\mathrm{D})}$.

\begin{figure}[t!]
\centering
\includegraphics[width=8.8cm]{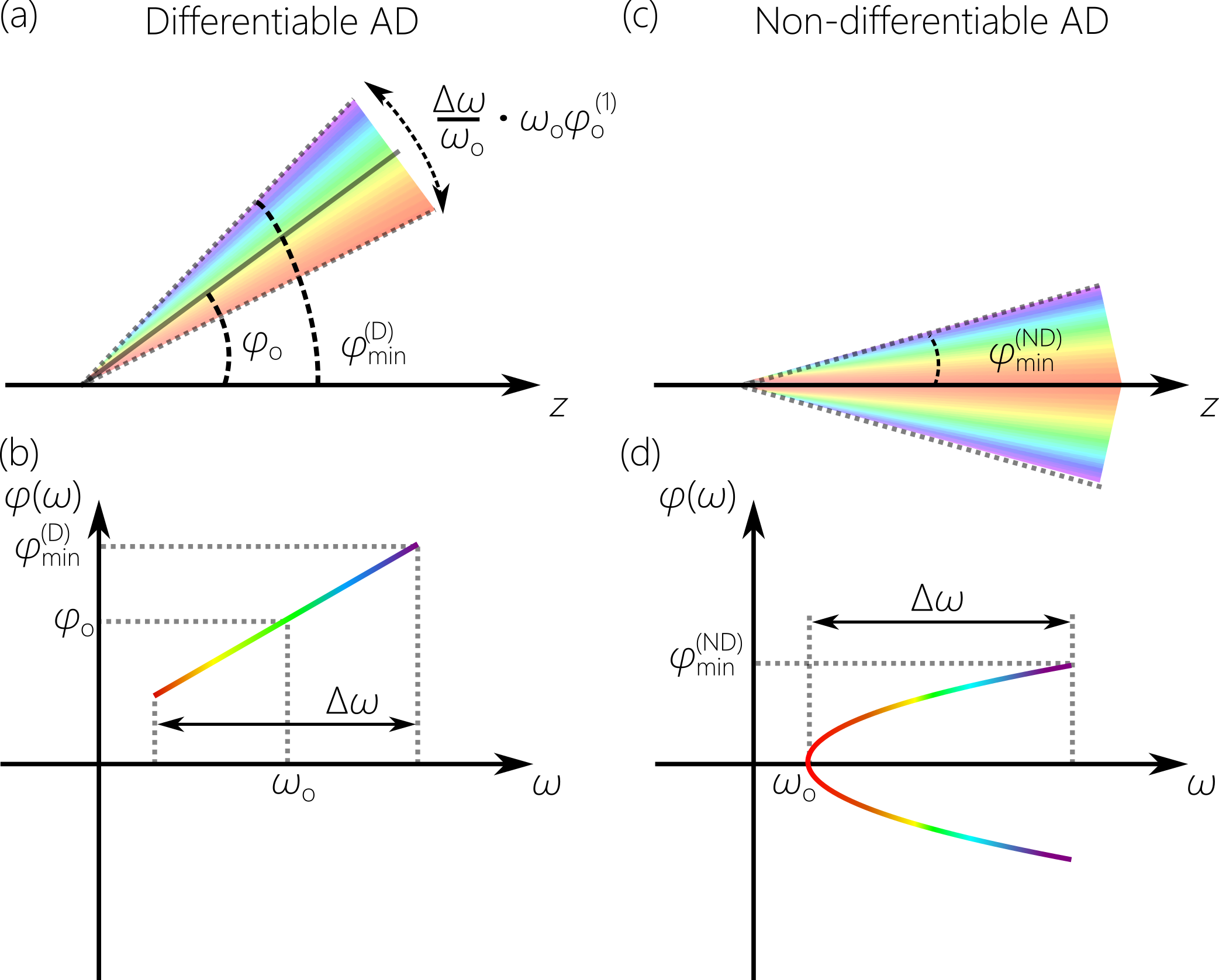}
\caption{(a) Illustration of the geometry for evaluating $\varphi_{\mathrm{min}}^{(\mathrm{D})}$ in the case of differentiable AD, and (b) the corresponding AD profile. (c,d) Same as (a,b) for non-differentiable AD to evaluate $\varphi_{\mathrm{min}}^{(\mathrm{ND})}$.}%\vspace{-5mm}
\label{fig:MaximumAngle}
\end{figure}

For differentiable AD, tuning the group velocity necessitates off-axis propagation, and the angle $\varphi_{\mathrm{min}}^{(\mathrm{D})}$ can be evaluated by examining the geometry in Fig.~\ref{fig:MaximumAngle}(a,b):
\begin{equation}\label{eq:MaxAngleDiffAD}
\varphi_{\mathrm{min}}^{(\mathrm{D})}\approx\varphi_{\mathrm{o}}+\frac{1}{2}\;\frac{\Delta\omega}{\omega_{\mathrm{o}}}\;\big|\omega_{\mathrm{o}}\varphi_{\mathrm{o}}^{(1)}\big|.    
\end{equation}
We do not consider the small-angle approximation ($\varphi_{\mathrm{o}}$ can take on \textit{any} value); rather, the approximation consists in neglecting the terms in the perturbative expansion of $\varphi(\omega)$ higher than $\varphi_{\mathrm{o}}^{(1)}$. From Eq.~\ref{eq:MaxAngleDiffAD}, the NA is determined by three factors:
\begin{enumerate}
\item the central propagation angle $\varphi_{\mathrm{o}}$; 
\item the first-order AD coefficient $\varphi_{\mathrm{o}}^{(1)}$; and 
\item the bandwidth $\Delta\omega$.
\end{enumerate}
Whereas $\Delta\omega$ is an independent parameter, $\varphi_{\mathrm{o}}$ and $\varphi_{\mathrm{o}}^{(1)}$ are mutually dependent in many cases, whereupon the number of parameters determining $\varphi_{\mathrm{min}}^{(\mathrm{D})}$ can be reduced to only $\varphi_{\mathrm{o}}$ and $\Delta\omega$. In principle, one can reduce the impact of the second term in Eq.~\ref{eq:MaxAngleDiffAD} by reducing the bandwidth $\Delta\omega$, resulting in a limit $\varphi_{\mathrm{min}}^{(\mathrm{D})}\rightarrow\varphi_{\mathrm{o}}$ when $\Delta\omega\rightarrow0$. The formula in Eq.~\ref{eq:MaxAngleDiffAD} can be applied to X-waves and pulsed Bessel beams as representative examples of differentiable AD \cite{Hall26JOSAA1}.

\subsubsection{Numerical aperture for X-waves}

For a superluminal X-wave with $\varphi_{\mathrm{o}}^{(1)}=0$ and $\widetilde{v}=c/\cos\varphi_{\mathrm{o}}$:
\begin{equation}\label{eq:NA_Xwave}
\varphi_{\mathrm{min}}^{(\mathrm{D})}=\varphi_{\mathrm{o}}=\cos^{-1}\widetilde{n}\approx\sqrt{2(1-\widetilde{n})},
\end{equation}
which is independent of the bandwidth $\Delta\omega$ (indeed, all the wavelengths travel at the same angle $\varphi_{\mathrm{o}}$), which makes it possible to exploit extremely large bandwidths without impacting the NA \cite{Saari97PRL}. Here a direct relationship exists between $\widetilde{v}$ and the NA, and deviation in $\widetilde{v}$ from $c$ necessitates a dramatic increase in the NA; e.g., $\widetilde{v}=1.1c$ requires $\varphi_{\mathrm{min}}^{(\mathrm{D})}\approx24.5^{\circ}$, which is already deep within the non-paraxial regime, and increasing $\widetilde{v}$ further increases $\varphi_{\mathrm{min}}^{(\mathrm{D})}$ even more.

\subsubsection{Numerical aperture for pulsed Bessel beams}

For a pulsed Bessel beam, $\omega_{\mathrm{o}}\varphi_{\mathrm{o}}^{(1)}=-\tan\varphi_{\mathrm{o}}$, so that $\varphi_{\mathrm{min}}^{(\mathrm{D})}$ depends on only $\varphi_{\mathrm{o}}$ and $\Delta\omega$:
\begin{equation}\label{eq:NAPulsedBessel}
\varphi_{\mathrm{min}}^{(\mathrm{D})}\approx\varphi_{\mathrm{o}}+\frac{1}{2}\frac{\Delta\omega}{\omega_{\mathrm{o}}}|\tan\varphi_{\mathrm{o}}|.
\end{equation}
Furthermore, $\widetilde{v}=c\cos\varphi_{\mathrm{o}}$, so that $\varphi_{\mathrm{min}}^{(\mathrm{D})}$ can be expressed in terms of $\widetilde{v}$ (or $\widetilde{n}$) and $\Delta\omega$ through the substitution $\varphi_{\mathrm{o}}=\cos^{-1}\tfrac{1}{\widetilde{n}}\approx\sqrt{2\tfrac{\widetilde{n}-1}{\widetilde{n}}}$. We plot $\varphi_{\mathrm{min}}^{(\mathrm{D})}$ in Fig.~\ref{fig:MaximumAngle_DiffAndNonDiff}(a) as we vary $\varphi_{\mathrm{o}}$ (or, equivalently $\widetilde{v}$) and $\tfrac{\Delta\omega}{\omega_{\mathrm{o}}}$. Increasing the deviation in $\widetilde{v}$ from $c$ or increasing $\tfrac{\Delta\omega}{\omega_{\mathrm{o}}}$ therefore increase the required NA. Even for a small deviation in $\widetilde{v}$ ($\widetilde{n}\rightarrow1$) and hence small $\varphi_{\mathrm{o}}$, we have:
\begin{equation}\label{eq:NA_PulsedBessel}
\varphi_{\mathrm{min}}^{(\mathrm{D})}\approx\sqrt{2\frac{\widetilde{n}-1}{\widetilde{n}}}\left(1+\frac{1}{2}\frac{\Delta\omega}{\omega_{\mathrm{o}}}\right).
\end{equation}
Except for short pulses (large $\Delta\omega$), $\varphi_{\mathrm{min}}^{(\mathrm{D})}$ is dictated mainly by $\varphi_{\mathrm{o}}$ (or $\widetilde{n}$). The baseline value when $\Delta\omega\rightarrow0$ is:
\begin{equation}
\varphi_{\mathrm{min}}^{(\mathrm{D})}\approx\sqrt{2\tfrac{\widetilde{n}-1}{\widetilde{n}}}\approx\sqrt{2(\widetilde{n}-1)}.
\end{equation}
This formula is similar to that in Eq.~\ref{eq:NA_Xwave} for an X-wave except for the reversal of the sign of the term under the square root to accommodate a subluminal group index $\widetilde{n}>1$. We thus expect once again that large deviations in $\widetilde{v}$ from $c$ occur only in the non-paraxial regime. For example, $\widetilde{v}=0.9c$ requires $\varphi_{\mathrm{o}}=25.9^{\circ}$. 

\subsection{Numerical aperture for non-differentiable AD}

\begin{figure}[t!]
\centering
\includegraphics[width=8.8cm]{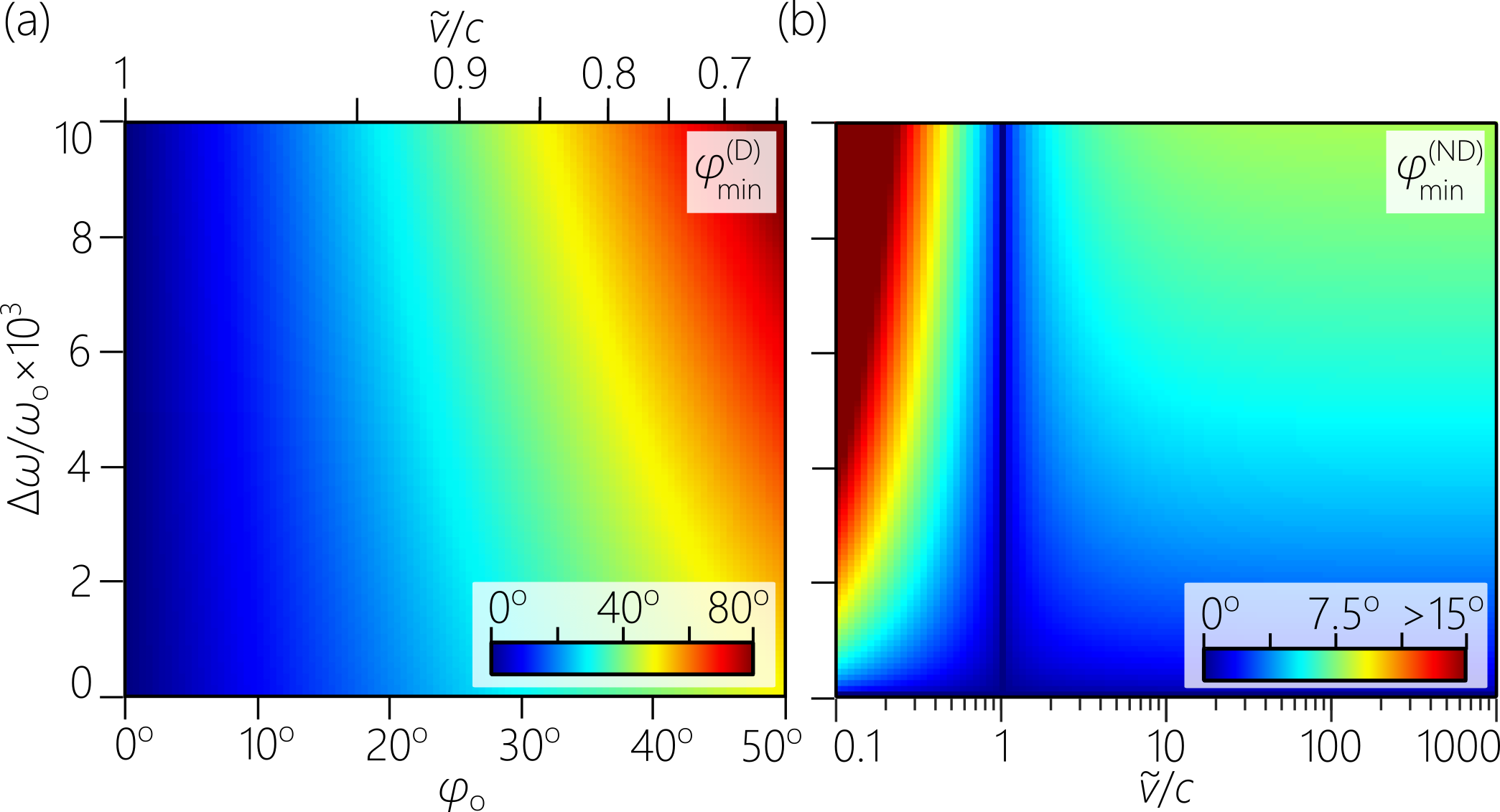}
\caption{(a) Plot of $\varphi_{\mathrm{min}}^{(\mathrm{D})}$ for a pulsed Bessel beam endowed with differentiable AD (Eq.~\ref{eq:NAPulsedBessel}) as a function of $\varphi_{\mathrm{o}}$ and $\Delta\omega$. We also plot the horizontal axis (top) in terms of $\widetilde{v}$, which is \textit{not} linearly related to $\varphi_{\mathrm{o}}$; $\widetilde{v}=c\cos\varphi_{\mathrm{o}}$. (b) Plot of $\varphi_{\mathrm{min}}^{(\mathrm{ND})}$ for an STWP with non-differentiable AD as a function of $\widetilde{v}$ and $\Delta\omega$. Note that the scale for $\varphi_{\mathrm{min}}^{(\mathrm{ND})}$ here is smaller than that in (a) for $\varphi_{\mathrm{min}}^{(\mathrm{D})}$, so that realizing a particular combination of group velocity $\widetilde{v}$ and bandwidth $\Delta\omega$ requires a smaller NA for non-differentiable AD than differentiable AD.}%\vspace{-5mm}
\label{fig:MaximumAngle_DiffAndNonDiff}
\end{figure}

For non-differentiable AD, we cannot rely on Eq.~\ref{eq:MaxAngleDiffAD} because $\varphi_{\mathrm{o}}^{(1)}$ is no longer defined. Instead, we make use of the constraint on the axial wave number $k_{z}$ for an STWP, $k_{z}(\omega)=k_{\mathrm{o}}+(\omega-\omega_{\mathrm{o}})/\widetilde{v}=\tfrac{\omega}{c}\cos\varphi$ (Eq.~\ref{eq:STWPconstraint}), from which we obtain $k_{x}(\omega)=\tfrac{\omega}{c}\sin\varphi=\sqrt{(\tfrac{\omega}{c})^{2}-k_{z}^{2}}$ [Fig.~\ref{fig:MaximumAngle}(c,d)], and thus:
\begin{equation}\label{eq:NonDiffNALargeAngle}
\sin\varphi_{\mathrm{min}}^{(\mathrm{ND})}=\sqrt{1-\left(\frac{1\pm\widetilde{n}\frac{\Delta\omega}{\omega_{\mathrm{o}}}}{1\pm\frac{\Delta\omega}{\omega_{\mathrm{o}}}}\right)^{2}},
\end{equation}
where the positive and negative signs are associated with superluminal and subluminal $\widetilde{v}$, respectively. We plot $\varphi_{\mathrm{min}}^{(\mathrm{ND})}$ in Fig.~\ref{fig:MaximumAngle_DiffAndNonDiff}(b) as a function of the bandwidth $\tfrac{\Delta\omega}{\omega_{\mathrm{o}}}$ and the group velocity $\widetilde{v}$. In the small-angle and narrow bandwidth approximations, Eq.~\ref{eq:NonDiffNALargeAngle} simplifies to:
\begin{equation}\label{eq:NonDiffNASmallAngle}
\varphi_{\mathrm{min}}^{(\mathrm{ND})}\approx\sqrt{2|1-\widetilde{n}|\frac{\Delta\omega}{\omega_{\mathrm{o}}}}.
\end{equation}

\subsection{Comparison between the NA for differentiable and non-differentiable AD}

The form of $\varphi_{\mathrm{min}}^{(\mathrm{ND})}$ in Eq.~\ref{eq:NonDiffNASmallAngle} shares the term $\sqrt{2|1-\widetilde{n}|}$ with $\varphi_{\mathrm{min}}^{(\mathrm{D})}$ for the X-wave in Eq.~\ref{eq:NA_Xwave} and that for the pulsed Bessel beam in Eq.~\ref{eq:NA_PulsedBessel}. The critical difference between $\varphi_{\mathrm{min}}^{(\mathrm{ND})}$ and $\varphi_{\mathrm{min}}^{(\mathrm{D})}$ is the role played by the bandwidth $\Delta\omega$. For differentiable AD, $\Delta\omega$ enters $\varphi_{\mathrm{min}}^{(\mathrm{D})}$ as an \textit{additive} term (Eqs.~\ref{eq:MaxAngleDiffAD} and \ref{eq:NA_PulsedBessel}), so that reducing $\Delta\omega$ does indeed reduce $\varphi_{\mathrm{min}}^{(\mathrm{D})}$, but even setting $\Delta\omega\rightarrow0$ reduces $\varphi_{\mathrm{min}}^{(\mathrm{D})}$ to a minimum of $\varphi_{\mathrm{min}}^{(\mathrm{D})}\rightarrow\varphi_{\mathrm{o}}\sim\sqrt{2|1-\widetilde{n}|}$. By contrast, in the case of non-differentiable AD, $\Delta\omega$ enters as a \textit{multiplicative} factor in $\varphi_{\mathrm{min}}^{(\mathrm{ND})}$ (Eq.~\ref{eq:NonDiffNASmallAngle}), so that reducing $\Delta\omega$ continuously reduces $\varphi_{\mathrm{min}}^{(\mathrm{ND})}$. In other words, if $\varphi_{\mathrm{o}}$ is in the non-paraxial regime for a wave packet endowed with differentiable AD, then reducing the bandwidth $\Delta\omega$ leaves the wave packet in the non-paraxial regime. By contrast, if an STWP has a large $\varphi_{\mathrm{min}}^{(\mathrm{ND})}$ in the non-paraxial regime, then reducing $\Delta\omega$ is guaranteed to reduce the NA and to reach the paraxial regime.

The increase in requisite propagation angle for differentiable AD (taking Eq.~\ref{eq:NA_PulsedBessel} as a representative) to realize the same group velocity with respect to that for non-differentiable AD (taking Eq.~\ref{eq:NonDiffNASmallAngle} as representative) is captured by the ratio:
\begin{equation}\label{eq:etaRatio}
\eta=\frac{\varphi_{\mathrm{min}}^{(\mathrm{D})}}{\varphi_{\mathrm{min}}^{(\mathrm{ND})}}\sim\frac{1}{\sqrt{\frac{\Delta\omega}{\omega_{\mathrm{o}}}}}+\tfrac{1}{2}\sqrt{\frac{\Delta\omega}{\omega_{\mathrm{o}}}}>1,
\end{equation}
which is independent of $\widetilde{v}$ and depends solely on the ratio $\tfrac{\Delta\omega}{\omega_{\mathrm{o}}}$. Non-differentiable AD thus always requires a smaller NA for the same $\widetilde{v}$ and $\tfrac{\Delta\omega}{\omega_{\mathrm{o}}}$. Typically $\tfrac{\Delta\omega}{\omega_{\mathrm{o}}}\ll1$, so that the first term in Eq.~\ref{eq:etaRatio} dominates the factor $\eta$. For example, if $\tfrac{\Delta\omega}{\omega_{\mathrm{o}}}=0.01$ ($\Delta\lambda=8$~nm at $\lambda_{\mathrm{o}}=800$~nm), then $\eta>10$; that is, the maximum angle required for a superluminal STWP is more than $10\times$ less than that for a superluminal X-wave having the same target group velocity and same bandwidth. Similarly, for the same parameters, the minimum angle required for a subluminal STWP is more than $10\times$ less than that for a subluminal pulsed Bessel beam having the same target group velocity and same bandwidth.

\begin{figure*}[t!]
\centering
\includegraphics[width=18.4cm]{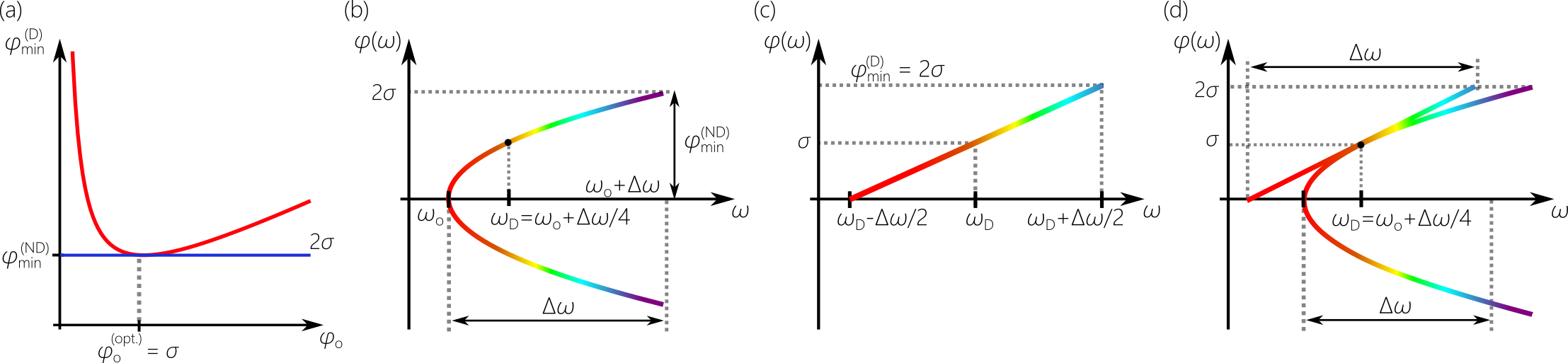}
\caption{(a) Plot of the required angle (NA) $\varphi_{\mathrm{min}}^{(\mathrm{D})}$ for the case of differentiable AD, corresponding to Eq.~\ref{eq:diff_minbandwidth} at fixed $\sigma$. The minimum angle value is $\varphi_{\mathrm{min}}^{(\mathrm{D})}=2\sigma=\varphi_{\mathrm{min}}^{(\mathrm{ND})}$, which occurs at $\varphi_{\mathrm{o}}^{(\mathrm{opt.})}=\sigma$. (b) AD profile for a field endowed with non-differentiable AD. (c) AD profile for a field endowed with differentiable AD that corresponds to the minimum point in (a). (d) Combining the AD profiles from (b) and (c) with $\omega_{\mathrm{D}}=\omega_{\mathrm{o}}+\tfrac{\Delta\omega}{4}$. Both fields have the same NA, $\varphi_{\mathrm{min}}^{(\mathrm{D})}=\varphi_{\mathrm{min}}^{(\mathrm{ND})}$, the same bandwidth $\Delta\omega$, and the same first-order AD coefficient (slope) at $\omega_{\mathrm{D}}$ for the differentiable AD profile and at $\omega_{\mathrm{o}}+\tfrac{\Delta\omega}{4}$ for the non-differentiable AD profile.}
\label{fig:Optimal}
\end{figure*}

To highlight the dramatic advantage conferred by non-differentiable AD, consider 1-nm-bandwidth pulses at $\lambda_{\mathrm{o}}\sim800$~nm ($\sim1$-ps pulse width), corresponding to $\eta\sim28.3$. To achieve superluminal group velocities $\widetilde{v}=1.1c$ and $2c$ with an X-wave utilizing this pulse, we require $\varphi_{\mathrm{min}}^{(\mathrm{D})}\sim24.5^{\circ}$ and $57.3^{\circ}$, respectively; whereas only $\varphi_{\mathrm{min}}^{(\mathrm{ND})}\sim0.9^{\circ}$ and $2^{\circ}$ are required for STWPs endowed with non-differentiable AD, respectively. Alternatively, to achieve subluminal group velocities $\widetilde{v}=0.9c$ and $0.5c$ with a pulsed Bessel beam utilizing this pulse, we require $\varphi_{\mathrm{min}}^{(\mathrm{D})}\sim25.6^{\circ}$ and $60.1^{\circ}$, respectively; whereas only $\varphi_{\mathrm{min}}^{(\mathrm{ND})}\sim0.9^{\circ}$ and $2^{\circ}$ are required for STWPs endowed with non-differentiable AD, respectively.

\subsection{Optimal NA in the small-angle limit}

The formula in Eq.~\ref{eq:MaxAngleDiffAD} accommodates any differentiable AD profile $\varphi(\omega)$. We pose the following question: which particular differentiable AD profile $\varphi(\omega)$ produces the optimal (i.e., minimum) value of $\varphi_{\mathrm{min}}^{(\mathrm{D})}$ for given $\widetilde{v}$ and $\tfrac{\Delta\omega}{\omega_{\mathrm{o}}}$? Note that we only need consider $\varphi_{\mathrm{o}}$ and $\omega_{\mathrm{o}}\varphi_{\mathrm{o}}^{(1)}$; higher-order AD terms do not impact $\widetilde{v}$.. Consequently, we examine only the subclass of linear AD profiles

Starting with the group index $\widetilde{n}=c/\widetilde{v}=\cos\varphi_{\mathrm{o}}-\omega_{\mathrm{o}}\varphi_{\mathrm{o}}^{(1)}\sin\varphi_{\mathrm{o}}$, we have $\omega_{\mathrm{o}}\varphi_{\mathrm{o}}^{(1)}=\tfrac{\cos\varphi_{\mathrm{o}}-\widetilde{n}}{\sin\varphi_{\mathrm{o}}}\approx\tfrac{1-\widetilde{n}}{\varphi_{\mathrm{o}}}$ in the small-angle approximation. That is, small $\varphi_{\mathrm{o}}$ or large $|1-\widetilde{n}|$ necessitate a large associated AD. We further assume that $\varphi_{\mathrm{o}}$ and $\varphi_{\mathrm{o}}^{(1)}$ are \textit{independent} of each other, so that our results here do \textit{not} apply to the pulsed Bessel beam (which is thus sub-optimal). We define the dimensionless quantity $\sigma$, where
\begin{equation}
\sigma^{2}=\frac{1}{2}\frac{\Delta\omega}{\omega_{\mathrm{o}}}\big|1-\widetilde{n}\big|,
\end{equation}
so that $\varphi_{\mathrm{min}}^{(\mathrm{D})}$ can be expressed in terms of $\sigma$ and $\varphi_{\mathrm{o}}$:
\begin{equation}\label{eq:diff_minbandwidth}
\varphi_{\mathrm{min}}^{(\mathrm{D})}\approx\varphi_{\mathrm{o}}+\frac{\sigma^{2}}{\varphi_{\mathrm{o}}}.
\end{equation}
For comparison, $\varphi_{\mathrm{min}}^{(\mathrm{ND})}$ in Eq.~\ref{eq:NonDiffNASmallAngle} for (on-axis) non-differentiable AD in the small-angle approximation is:
\begin{equation}\label{eq:nondiff_minbandwidth}
\varphi_{\mathrm{min}}^{(\mathrm{ND})}\approx2\sigma.
\end{equation}

The quantity $\sigma$ contains the two targeted parameters, the group velocity $\widetilde{v}$ and the bandwidth $\Delta\omega$, and we thus hold it fixed. Consequently, the NA for the non-differentiable AD case is fixed, while that for differentiable AD varies with $\varphi_{\mathrm{o}}$. We plot $\varphi_{\mathrm{min}}^{(\mathrm{D})}$ from Eq.~\ref{eq:diff_minbandwidth} and $\varphi_{\mathrm{min}}^{(\mathrm{ND})}$ from Eq.~\ref{eq:nondiff_minbandwidth} in Fig.~\ref{fig:Optimal}(a) as a function of $\varphi_{\mathrm{o}}$ for fixed $\sigma$. The plot of $\varphi_{\mathrm{min}}^{(\mathrm{D})}$ reaches a minimum -- representing the optimal NA for differentiable AD -- when $\varphi_{\mathrm{o}}^{(\mathrm{opt})}=\sigma$, which yields the optimal value for the NA of $\varphi_{\mathrm{min}}^{(\mathrm{D,opt})}=2\sigma=\varphi_{\mathrm{min}}^{(\mathrm{ND})}$. In other words, for fixed group velocity $\widetilde{v}$ and bandwidth $\Delta\omega$ (as combined in the parameter $\sigma$), optimal use of the NA of the optical system with differential AD yields the same NA as for non-differentiable AD.

It is useful to examine in more detail this optimal case for differentiable AD in comparison with non-differentiable AD. To that end, we plot in Fig.~\ref{fig:Optimal}(b) the AD profile for non-differentiable AD associated with superluminal $\widetilde{v}$, $\varphi^{(\mathrm{ND})}(\omega)=\sqrt{2(1-\widetilde{n})\tfrac{\omega-\omega_{\mathrm{o}}}{\omega_{\mathrm{o}}}}$, where $\omega>\omega_{\mathrm{o}}$ and $\widetilde{n}<1$. The bandwidth $\Delta\omega$ extends from $\omega=\omega_{\mathrm{o}}$ to $\omega=\omega_{\mathrm{o}}+\Delta\omega$, and we write $\varphi^{(\mathrm{ND})}(\omega)=2\sigma\sqrt{\tfrac{\omega-\omega_{\mathrm{o}}}{\Delta\omega}}$. The slope of the AD profile is $\tfrac{d\varphi^{(\mathrm{ND})}}{d\omega}=\tfrac{\sigma}{\sqrt{\Delta\omega(\omega-\omega_{\mathrm{o}})}}$, $\omega\neq\omega_{\mathrm{o}}$. We draw attention to the following values of the angle at prescribed frequencies; at $\omega_{\mathrm{o}}$, $\omega=\omega_{\mathrm{o}}+\tfrac{\Delta\omega}{4}$, and  $\omega=\omega_{\mathrm{o}}+\Delta\omega$ we have:
\begin{eqnarray}
\varphi^{(\mathrm{ND})}(\omega_{\mathrm{o}})\!\!\!\!&=&\!\!\!\!0,\quad\frac{d\varphi^{(\mathrm{ND})}}{d\omega}\rightarrow\infty,\nonumber\\
\varphi^{(\mathrm{ND})}(\omega_{\mathrm{o}}+\tfrac{\Delta\omega}{4})\!\!\!\!&=&\!\!\!\!\sigma,\quad\frac{d\varphi^{(\mathrm{ND})}}{d\omega}\biggr|_{\omega_{\mathrm{o}}+\Delta\omega/4}=\frac{2\sigma}{\Delta\omega},\nonumber\\
\varphi^{(\mathrm{ND})}(\omega_{\mathrm{o}}+\Delta\omega)\!\!\!\!&=&\!\!\!\!2\sigma,\quad\frac{d\varphi^{(\mathrm{ND})}}{d\omega}\biggr|_{\omega_{\mathrm{o}}+\Delta\omega}=\frac{\sigma}{\Delta\omega}.
\end{eqnarray}
Note that as $\varphi^{(\mathrm{ND})}(\omega)$ increases, $\tfrac{d\varphi^{(\mathrm{ND})}}{d\omega}$ concomitantly decreases; specifically, $\varphi^{(\mathrm{ND})}(\omega_{\mathrm{o}}+\tfrac{\Delta\omega}{4})=\tfrac{1}{2}\varphi^{(\mathrm{ND})}(\omega_{\mathrm{o}}+\Delta\omega)$ while $\tfrac{d\varphi^{(\mathrm{ND})}}{d\omega}\big|_{\omega_{\mathrm{o}}+\Delta\omega/4}=2\tfrac{d\varphi^{(\mathrm{ND})}}{d\omega}\big|_{\omega_{\mathrm{o}}+\Delta\omega}$.

Next, consider an example of differentiable AD that includes only the first two terms in Eq.~\ref{eq:ADexpansion}, $\varphi^{(\mathrm{D})}(\omega)=\varphi_{\mathrm{D}}+(\omega-\omega_{\mathrm{D}})\varphi_{\mathrm{D}}^{(1)}$, where $\omega_{\mathrm{D}}$ is the central frequency of the AD profile and $\varphi_{\mathrm{D}}=\varphi(\omega_{\mathrm{D}})$ [Fig.~\ref{fig:Optimal}(c)]. For simplicity, we plot only $\varphi(\omega)$, but we assume that a symmetrized AD profile is implemented. We assume that the target group velocity $\widetilde{v}$ and the bandwidth $\Delta\omega$ are equal to those for the non-differentiable AD scenario above. The spectrum extends from $\omega=\omega_{\mathrm{D}}-\tfrac{\Delta\omega}{2}$ to $\omega=\omega_{\mathrm{D}}+\tfrac{\Delta\omega}{2}$. The optimal NA condition derived above entails that $\varphi(\omega_{\mathrm{D}})=\varphi_{\mathrm{D}}=\sigma$, $\varphi(\omega_{\mathrm{D}}+\tfrac{\Delta\omega}{2})=2\sigma$, and $\varphi_{\mathrm{D}}^{(1)}=\tfrac{2\sigma}{\Delta\omega}$, which imply that $\varphi(\omega_{\mathrm{D}}-\tfrac{\Delta\omega}{2})=0$. Rotating the plot so that $\omega$ becomes the vertical axis, the AD spectrum is V-shaped, thus corresponding to what we have called V-waves \cite{Hall21PRAVwave}; see also Part~I of this tutorial for more details \cite{Hall26JOSAA1}.

To establish the relationship between the spectra for non-differentiable AD [Fig.~\ref{fig:Optimal}(b)] and the optimal differentiable AD [Fig.~\ref{fig:Optimal}(c)], we find the frequency at which the AD coefficient $\tfrac{d\varphi}{d\omega}$ is equal in both. For the differentiable AD scenario in Fig.~\ref{fig:Optimal}(c), the slope is constant $\varphi_{\mathrm{D}}^{(1)}=\tfrac{d\varphi^{(\mathrm{D})}}{d\omega}=\tfrac{2\sigma}{\Delta\omega}$, whereas the slope varies with $\omega$ for the non-differentiable AD scenario in Fig.~\ref{fig:Optimal}(b) with $\tfrac{d\varphi^{(\mathrm{ND})}}{d\omega}=\tfrac{2\sigma}{\Delta\omega}$ at $\omega=\omega_{\mathrm{o}}+\tfrac{\Delta\omega}{4}$. We thus can make the association $\omega_{\mathrm{D}}=\omega_{\mathrm{o}}+\tfrac{\Delta\omega}{4}$, whereupon the linear AD profile is tangential to the non-differentiable AD profile [Fig.~\ref{fig:Optimal}(d)]. The two profiles have the same bandwidth $\Delta\omega$, yield the same group velocity $\widetilde{v}$, and have the same maximum angle $\varphi_{\mathrm{min}}^{(\mathrm{D})}=\varphi_{\mathrm{min}}^{(\mathrm{ND})}=2\sigma$, which is the optimal scenario.

From this result, we reach a critical conclusion: despite the paucity of previous results, conventional AD is capable \textit{in principle} of producing large deviations in $\widetilde{v}$ from $c$ in free space within the paraxial regime. However, achieving this requires satisfying two conditions: (1) independent control over $\varphi_{\mathrm{o}}$ and $\varphi_{\mathrm{o}}^{(1)}$; and (2) a large value of $\omega_{\mathrm{o}}\varphi_{\mathrm{o}}^{(1)}$ is typically needed. Both of these requirements remain challenges for optics. Moreover, even if these two conditions are met, the resulting AD profile is likely to be accompanied with higher-order AD terms and AD-induced GVD (even for purely linear AD). Such dispersive effects are absent from non-differentiable AD. This highlights the uniqueness and usefulness of the recently discovered non-differentiable AD.

As a numerical example, consider a wave packet endowed with differentiable AD, $\tfrac{\Delta\omega}{\omega_{\mathrm{D}}}=\tfrac{1}{100}$ (corresponding, for example, to a pulse with bandwidth $\Delta\lambda=8$~nm at a central wavelength $\lambda_{\mathrm{o}}=800$~nm). To achieve a superluminal group velocity with $\widetilde{n}=0.9$, the AD required is $\omega_{\mathrm{D}}\varphi_{\mathrm{D}}^{(1)}\approx4.5$, which is close to the maximum that can be achieved with a diffraction grating \cite{Hall25APLP,Hall26JOSAA1}. The central angle for such a wave packet is $\varphi_{\mathrm{D}}\approx1.3^{\circ}$, with an NA corresponding to the angle $\varphi_{\mathrm{min}}^{(\mathrm{D})}\approx2.6^{\circ}$. To reduce the requisite angle for a fixed group velocity, we can reduce the bandwidth, say to $\tfrac{\Delta\omega}{\omega_{\mathrm{D}}}=\tfrac{1}{800}$ (corresponding, for example, to a pulse with bandwidth $\Delta\lambda=1$~nm at a central wavelength $\lambda_{\mathrm{o}}=800$~nm), which reduces the NA to an angle $\varphi_{\mathrm{min}}^{(\mathrm{D})}\approx0.9^{\circ}$. However, this reduction in NA necessitates increasing the AD to $\omega_{\mathrm{D}}\varphi_{\mathrm{D}}^{(1)}\approx12.65$, a value that cannot be produced by conventional means. Furthermore, the increase in AD signifies a concomitant increases in the AD-induced GVD accompanying the wave packet. By contrast, the wave packet endowed with non-differentiable AD is free of GVD for any bandwidth. Although conventional devices do not produce either of these wave packets, recently developed universal AD synthesizers \cite{Yessenov22AOP,Yessenov22NC,Hall24JOSAA,Romer25JOpt,Yessenov25JOSAA} can realize both.

\section{Maximum relative group delay}

\begin{figure}[t!]
\centering
\includegraphics[width=8.8cm]{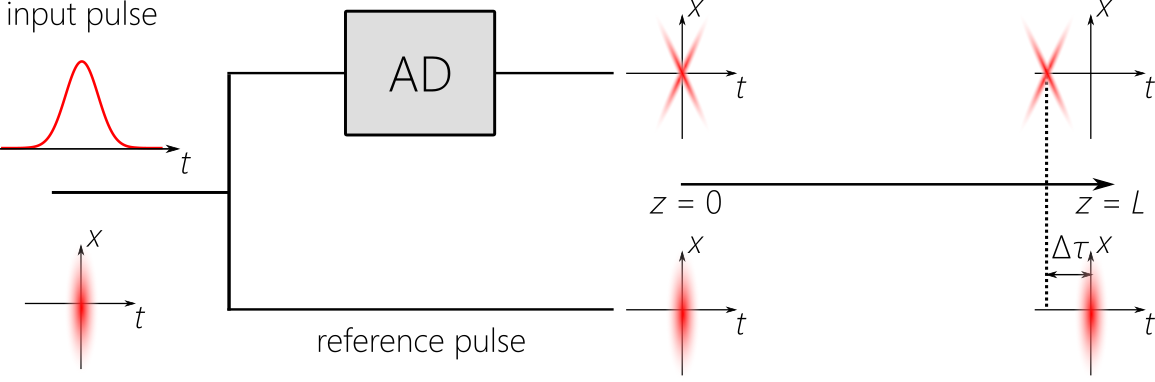}
\caption{Conceptual scheme for estimating the group velocity of a structured optical wave packet. A collimated pulse (on the left, with a separable spatiotemporal profile) is shaped into a wave packet by introducing AD. The relative delay $\Delta\tau$ between this wave packet and a reference pulse after traveling the same distance $L$ is measured to estimate $\widetilde{v}$.}
\label{fig:Setup}
\end{figure}

Estimating the group velocity $\widetilde{v}$ of a wave packet having a pulse width of $\Delta T$ in free space experimentally is typically achieved by measuring the relative group delay $\Delta\tau=\frac{1}{c}(1-\widetilde{n})L$ between the wave packet and a reference pulse traveling at $c$, where $L$ is the common path length for the pulses; see Fig.~\ref{fig:Setup}. For $\widetilde{v}\rightarrow c$ (which has been the case for all reported experiments except those for STWPs), $\Delta\tau$ is small -- typically smaller than the pulse width $\Delta T$, which makes the measurement challenging. To address this challenge one may follow one of two different strategies:
\begin{enumerate}
\item To increase $\Delta\tau$ at fixed $\widetilde{n}$, one may increase $L$ at fixed pulse width $\Delta T$, which requires that the wave packet maintains its spatiotemporal integrity over this extended distance $L$.
\item Alternatively, one may reduce the pulse width $\Delta T$ to improve the ratio $\tfrac{\Delta\tau}{\Delta T}$; i.e., increase the ratio of the relative delay as a fraction of the pulse width.
\end{enumerate}

However, increasing $L$ is limited by the onset of diffractive spatial spreading at the diffraction distance $L_{\mathrm{diff}}$, or of dispersive temporal spreading at the AD-induced dispersion length $L_{\mathrm{disp}}\sim\tfrac{(\Delta T)^{2}}{|k_{2}|}$. Therefore, one cannot increase $L$ indefinitely to improve the measurement accuracy. Moreover, reducing the pulse width $\Delta T$ (increasing the bandwidth $\Delta\omega$) in turn reduces $L_{\mathrm{disp}}$, and may increase the NA, which may in turn reduce $L_{\mathrm{diff}}$. These factors all need to be considered when maximizing $\Delta\tau$.

The maximum achievable $\Delta\tau$ for fixed $\Delta T$ can be estimated by replacing $L$ with the shorter of the diffraction length $L_{\mathrm{diff}}$ or the dispersion length $L_{\mathrm{disp}}$. AD-free wave packets (X-waves) are free of GVD and are thus primarily limited by diffraction, where $L_{\mathrm{diff}}= \frac{c}{\delta\omega(1-\cos\varphi_\mathrm{o})}$ as obtained by evaluating the walk-off between the X-wave and an accompanying pilot envelope arising from the spectral uncertainty \cite{Yessenov19OE,Kondakci19OL}. STWPs endowed with non-differentiable AD are also free of GVD, and $\Delta\tau$ is thus limited by diffraction, with $L_{\mathrm{diff}}\sim\tfrac{c}{\delta\omega|1-\widetilde{n}|}$ and $\Delta\tau_{\mathrm{max}}\sim\tfrac{1}{\delta\omega}$ as established in \cite{Yessenov19OE,Kondakci19OL}, where $\delta\omega$ is the spectral uncertainty. Therefore, $\Delta\tau$ depends solely on $\delta\omega$ and is independent of $\omega_\mathrm{o}$, the bandwidth $\Delta\omega$, and $\widetilde{v}$. For a pulsed Bessel beam endowed with differentiable-AD, $L_{\mathrm{diff}}\sim D/\tan\varphi_{\mathrm{o}}$ for an aperture of diameter~$D$ \cite{McGloin05CP,Mazilu10LPR}, whereas the dispersion length is $L_{\mathrm{disp}}\sim\tfrac{\omega_{\mathrm{o}}c(\Delta T)^{2}}{(\omega_{\mathrm{o}}\varphi_{\mathrm{o}}^{(1)})^{2}}=\tfrac{\omega_{\mathrm{o}}c(\Delta T)^{2}}{\tan^{2}\varphi_{\mathrm{o}}}$ for small $\varphi_{\mathrm{o}}$, whereupon $\omega_{\mathrm{o}}ck_{2}\approx-(\omega_{\mathrm{o}}\varphi_{\mathrm{o}}^{(1)})^{2}=-\tan^{2}\varphi_{\mathrm{o}}$. In Ref.~\cite{Giovannini15Science}, $\varphi_{\mathrm{o}}\approx0.26^{\circ}$, yielding $L_{\mathrm{disp}}\approx1.4$~km (taking $\lambda_{\mathrm{o}}=800$~nm and $\Delta T=200$~fs) and $L_{\mathrm{diff}}\approx5.5$~m, so that $L_{\mathrm{diff}}$ dominates in this case. The propagation distance in \cite{Giovannini15Science} was $L=1$~m, which could only be slightly increased ($L<L_{\mathrm{diff}}=5.5$~m). Reducing $\Delta T$ (increasing the bandwidth) is not useful here because it does not increase $L_{\mathrm{diff}}$. Increasing $\varphi_{\mathrm{o}}$ to increase the deviation in $\widetilde{v}$ from $c$ leads to a rapid decrease in both $L_{\mathrm{disp}}$ and $L_{\mathrm{diff}}$.

%\textcolor{blue}{For pulsed Bessel beams incorporating differentiable AD, $\Delta\tau_\mathrm{max}/\Delta T \sim \omega_\mathrm{o}/\Delta\omega$: narrow-bandwidth pulses at high carrier frequencies are required to maximize the measurable group delay.}

\section{Discussion}

\subsection{Summary}

We summarize in what follows the salient points established with regards to the tunability of the group velocity $\widetilde{v}$ of a pulsed beam along the $z$-axis in free space. First, with regards to conventional \textit{differentiable} AD \cite{Hall26JOSAA1}:
\begin{enumerate}
\item Differentiable AD can be readily introduced via a diffractive or dispersive element (e.g., a diffraction grating, a prism, a metasurface, etc.).
\item Differentiable AD only enables tuning $\widetilde{v}$ for off-axis fields. 
\item By contrast to conventional wisdom, an AD-endowed wave packet need not be subluminal. In principle, differentiable AD can yield arbitrary $\widetilde{v}$ in free space -- whether subluminal, superluminal, or even negative -- through a combination of geometric \textit{and} interferometric effects.
\item However, this broad tunability requires independent control over multiple terms in the AD profile. This is currently a challenge for optics.
\item The requirement for off-axis propagation usually results in a large NA to tune $\widetilde{v}$ even when the pulse is very long (narrow bandwidth) because the off-axis angle $\varphi_{\mathrm{o}}$ and the bandwidth are additive factors in determining the NA.
\item Consequently, only minute changes in $\widetilde{v}$ are accessible in the paraxial regime. 
\item Two factors limit the distance over which the group velocity can be utilized: diffractive spreading of the spatial profile and dispersive spreading of the temporal profile that accompanies differentiable AD.
\end{enumerate}

By contrast, recently developed non-differentiable AD associated with STWPs alleviates these challenges.
\begin{enumerate}
\item Arbitrary group velocities can be produced in the paraxial regime.
\item The bandwidth $\Delta\omega$ is a multiplicative factor in determining the NA. Consequently, the NA can always be reduced for a prescribed $\widetilde{v}$ by reducing $\Delta\omega$.
\item The wave packet is free of AD-induced GVD, and the propagation-invariant distance is determined by the spectral uncertainty.
\item Crucially, the numerical aperture is utilized optimally. No wave packet endowed with differentiable AD can have the same group velocity and bandwidth as the non-differentiable AD scenario and yet require a smaller NA.
\item However, introducing non-differentiable AD requires more complex setups that combine spectral analysis with spectral phase modulation of the spatial profile. Rapid progress is being made in constructing compact, simple universal AD synthesizers.
\end{enumerate}

\subsection{Flying focus pulses}

The summary given above provides a useful rule of thumb: any wave packet whose group velocity $\widetilde{v}$ can be tuned only in the close vicinity of $c$ in the paraxial regime is most likely undergirded by differential AD. By contrast, any wave packet whose group velocity $\widetilde{v}$ can be tuned substantially away from $c$ while remaining in the paraxial regime \textit{necessarily} employs some form of non-differentiable AD.

\begin{figure}[t!]
\centering
\includegraphics[width=8.8cm]{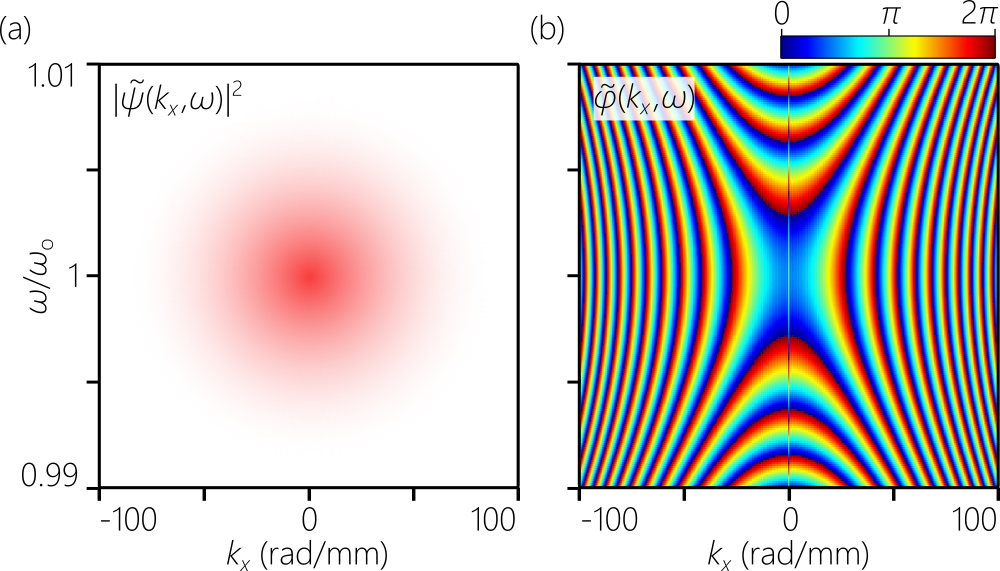}
\caption{(a) Intensity of the spatiotemporal profile $|\widetilde{\psi}(k_{x},\omega)|^2$ and (b) the associated spatiotemporal phase $\widetilde{\phi}(k_{x},\omega)$; $\widetilde{\psi}(k_{x},\omega)=|\widetilde{\psi}(k_{x},\omega)|e^{i\widetilde{\phi}(k_{x},\omega)}$.}
\label{fig:FF}
\end{figure}

This rule of thumb applies to wave packets that maintain their characteristics with propagation. An interesting exception is the so-called `flying focus' \cite{SaintMarie17Optica,Froula18NP} that demonstrates a tunable $\widetilde{v}$ over a broad span of values similar to that of STWPs in free space, while requiring a simple optical setup for its synthesis. A collimated pulse is temporally dispersed and then focused via a chromatic lens. Changing the pulse dispersion allows tuning $\widetilde{v}$ for the \textit{on-axis} wave packet. However, the on-axis wave packet comprises only a fraction of the initial pulse spectrum, and the central wavelength of this reduced-bandwidth wave packet changes with propagation along the axis through the lens focal volume. In other words, the on-axis spectrum (which is a fraction of the total spectrum, corresponding to a longer pulse) evolves with propagation. The associated spatiotemporal spectrum is \cite{SaintMarie17Optica}:
\begin{equation}
\widetilde{\psi}(k_{x},\omega)=\exp\left\{\frac{i\phi''}{2}\Omega^2\right\}\;
\exp\left\{\,-\frac{W_\mathrm{o}^2}{4}k_x^2+i\frac{cf(\omega)}{2\omega}k_x^2\right\},
\end{equation}
where $W_{\mathrm{o}}$ is the beam waist, $f(\omega)$ is the lens focal length as a function of $\omega$, and $\phi''$ is the group delay dispersion (GDD) of the initial pulse.
Plotting the spatiotemporal spectral amplitude $|\widetilde{\psi}(k_{x},\omega)|$ in $(k_{x},\omega)$ space in Fig.~\ref{fig:FF}(a) reveals that it is \textit{not} endowed with AD. However, the spatiotemporal spectral phase $\widetilde{\varphi}(k_{x},\omega)$ plotted in Fig.~\ref{fig:FF}(b) has a complex profile that is non-differentiable at the center of the spatiotemporal spectrum, $k_{x}=0$ and $\omega=\omega_\mathrm{o}$. This non-differentiable spectral phase facilitates the large deviations of $\widetilde{v}$ from $c$, but the absence of AD results in profound spectral variations occurring in the wave packet with propagation.

To halt this axial spectral evolution (reminiscent of STWPs with intentional axial spectral encoding \cite{Motz20arxiv,Hall25OLaxial}) and produce a wave packet with stable spectrum that makes use of the full pulse bandwidth on the optical axis (the on-axis pulse width is \textit{not} elongated), a modification of the above-described `chromatic' flying focus was proposed and realized. By contrast to the chromatic flying focus, this `achromatic' flying focus \cite{Pigeon24OE} maintains the entire bandwidth along the axis. Nevertheless, only minute changes in $\widetilde{v}$ were accessible, by contrast to the large tuning range of $\widetilde{v}$ for the chromatic flying focus. This suggests that the AD underpinning this achromatic flying focus is differentiable. Subsequently, an `ideal flying focus' was theoretically studied \cite{Ramsey23PRA} in which the group velocity is widely tunable, the full bandwidth is maintained on-axis, and the wave packet is propagation invariant. However, it was subsequently shown that this ideal flying focus coincides with STWPs \cite{Almeida25OL}, confirming STWPs as the unique family of propagation-invariant wave packets with fixed group velocity.

\subsection{Comparison of angular and chromatic dispersion}\label{eq:ComparisonADCD}

It is instructive to compare AD to chromatic dispersion (CD), which arises from the wavelength dependence of the refractive index \cite{SalehBook07}. Both AD and CD enable tuning the group velocity $\widetilde{v}$. Whereas AD allows for tuning $\widetilde{v}$ in free space without requiring an optical medium or photonic structure, chromatic dispersion requires such a medium or structure, which has been the foundation for so-called `slow' and `fast' light \cite{Kurgin08Book}.

Consider here a plane-wave pulse traveling in a dispersive medium of refractive index $n(\omega)$, so that all the frequencies are co-aligned with the $z$-axis and $k(\omega)\!=\!n(\omega)\tfrac{\omega}{c}$. We expand the \textit{refractive index} into a Taylor series as we did for the \textit{propagation angle} (compare to Eq.~\ref{eq:ADexpansion}):
\begin{equation}\label{eq:CDexpansion}
n(\omega)\!=\!n(\omega_{\mathrm{o}}+\Omega)\!=\!n_{\mathrm{o}}+n_{\mathrm{o}}^{(1)}\Omega+\tfrac{1}{2}n_{\mathrm{o}}^{(2)}\Omega^{2}+\cdots,
\end{equation}
where $n_{\mathrm{o}}\!=\!n(\omega_{\mathrm{o}})$, and $n_{\mathrm{o}}^{(m)}\!=\!\tfrac{d^{m}n}{d\omega^{m}}\big|_{\omega_{\mathrm{o}}}$ for $m$ integer. The phase velocity along the $z$-axis is $v_{\mathrm{ph}}=\tfrac{\omega}{k}\big|_{\omega_{\mathrm{o}}}=\tfrac{c}{n_{\mathrm{o}}}$, and the group velocity $v_{\mathrm{g}}$ along the propagation axis is (compare to Eq.~\ref{eq:GroupVelocityGeneral}):
\begin{equation}
v_{\mathrm{g}}=\tfrac{1}{\frac{dk}{d\omega}\big|_{\omega_{\mathrm{o}}}}=\frac{c}{n_{\mathrm{o}}+\omega_{\mathrm{o}}n_{\mathrm{o}}^{(1)}}\;.
\end{equation}
These formulas for CD clearly map to their counterparts in presence of AD with the correspondence: $n_{\mathrm{o}}\leftrightarrow\cos\varphi_{\mathrm{o}}$ and $\omega_{\mathrm{o}}n_{\mathrm{o}}^{(1)}\leftrightarrow-\omega_{\mathrm{o}}\varphi_{\mathrm{o}}^{(1)}\sin\varphi_{\mathrm{o}}$, which results from the correspondence between $k_{z}(\omega)=\tfrac{\omega}{c}\cos\varphi(\omega)$ in presence of AD and $k_{z}(\omega)=\tfrac{\omega}{c}n(\omega)$ in presence of CD. 

Conventionally, normal dispersion refers to $\widetilde{n}_{\mathrm{o}}^{(1)}>0$ and anomalous dispersion to $\widetilde{n}_{\mathrm{o}}^{(1)}<0$. The difference in sign for the contributions of $n_{\mathrm{o}}^{(1)}$ and $\varphi_{\mathrm{o}}^{(1)}$ to the group velocity in the presence of CD and AD, respectively, is behind our definition for normal and anomalous AD in Ref.~\cite{Hall26JOSAA1}. Table~\ref{Table:normal} summarizes the salient comparisons between CD and AD in the context of controlling the group velocity.

\begin{table}[t!]
\caption{Comparison of angular dispersion (AD) and chromatic dispersion (CD) with respect to the group velocity.}\label{Table:normal}
\renewcommand{\arraystretch}{1.8}
\begin{tabular}{|l|l|l|}
 \hline
  & CD & AD\\ \hline\hline
Origin&material property&\parbox[c]{3.2cm}{intrinsic field structure}\\ \hline

$k_{z}(\omega)$ & $\tfrac{\omega}{c}n(\omega)$ & $\tfrac{\omega}{c}\cos\varphi(\omega)$ \\ \hline

Dispersion & $n(\omega)\!=\!n_{\mathrm{o}}+n_{\mathrm{o}}^{(1)}\Omega+\cdots$ & $\varphi(\omega)\!=\!\varphi_{\mathrm{o}}+\varphi_{\mathrm{o}}^{(1)}\Omega+\cdots$ \\ \hline

phase index&$n_{\mathrm{ph}}=n_{\mathrm{o}}$ & $n_{\mathrm{ph}}=\cos\varphi_{\mathrm{o}}$ \\ \hline

phase vel. &$v_{\mathrm{ph}}\!=\!\tfrac{c}{n_{\mathrm{o}}}$ & $v_{\mathrm{ph}}\!=\!\tfrac{c}{\cos\varphi_{\mathrm{o}}}$ \\ \hline

group index&$\widetilde{n}=n_{\mathrm{o}}+\omega_{\mathrm{o}}n_{\mathrm{o}}^{(1)}$ & $\widetilde{n}=\cos\varphi_\mathrm{o} -\omega_{\mathrm{o}}\varphi_{\mathrm{o}}^{(1)}\sin\varphi_{\mathrm{o}}$ \\ \hline

group vel.& $v_{\mathrm{g}}\!=\!\frac{c}{n_{\mathrm{o}}+\omega_{\mathrm{o}}n_{\mathrm{o}}^{(1)}}$ & $\widetilde{v}\!=\!\frac{c}{\cos\varphi_{\mathrm{o}}-\omega_{\mathrm{o}}\varphi_{\mathrm{o}}^{(1)}\sin\varphi_{\mathrm{o}}}$\\ 
\hline
\end{tabular}\end{table}

The GVD coefficient $k_{2}=\frac{d^{2}k}{d\omega^{2}}\big|_{\omega_{\mathrm{o}}}$ accompanying chromatic dispersion is:
\begin{equation}
c\omega_{\mathrm{o}}k_{2}=\omega_{\mathrm{o}}^{2}n_{\mathrm{o}}^{(2)}+2\omega_{\mathrm{o}}n_{\mathrm{o}}^{(1)},
\end{equation}
which is to be compared to the formula for the AD-induced GVD coefficient in Ref.~\cite{Hall26JOSAA1}. This reveals a fundamental difference between the two scenarios of AD-induced and CD-induced GVD for on-axis propagation: whereas AD-induced GVD is always anomalous for on-axis wave packets, CD-induced GVD can be either normal or anomalous. We have shown above that non-differentiable AD can eliminate AD-induced GVD. To date, no optical materials or photonic structures have been identified that display `non-differentiable CD' that would yield, in principle, GVD-free propagation over an extended bandwidth.

Finally, note that CD is inevitably accompanied by either optical absorption or gain arising from the imaginary part of the refractive index, which is related to the real part via the Kramers-Kronig relations \cite{SalehBook07}. By contrast, AD is \textit{not} accompanied by absorption or gain.

\subsection{Further extensions}

We focused in this tutorial on free-space propagation. Nevertheless, STWPs can be generalized to dispersive media in which the group velocity can also be tuned, leading to a host of striking phenomena predicted in Refs.~\cite{Malaguti08OL,Malaguti09PRA} and observed recently \cite{Hall23LPR,Hall23NatPhys,Hall24ACSP,Hall25PRA}. The limits on group-velocity tunability via non-differentiable AD in dispersive media have not yet been established \cite{Yessenov21ACSPhot}. This can be further extended to other dispersive platforms, including space-time surface plasmon polaritons (SPPs) at metal-dielectric interfaces \cite{Schepler20ACSPhot,Ichiji25NC}, in which non-differentiable AD yields diffraction-free SPPs with a tunable group velocity.

We have couched our formulation in terms of coherent optical pulses, but these results extend to \textit{spectrally incoherent} fields, whereupon $\widetilde{v}$ becomes a `coherence group velocity' or `coherence speed'. This is the velocity of the peak of the coherence function with respect to that of the spatially unstructured spectrally incoherent field \cite{Yessenov19Optica,Yessenov19OL}. Much work is needed along these lines, which is expected to enrich the study of optical coherence.

Because the maximum relative group delay $\Delta\tau$ for an STWP with respect to a collimated pulse is $\Delta\tau\sim\tfrac{1}{\delta\omega}$, efforts have been directed to reduce $\delta\omega$ and thus increase $\Delta\tau$ (and in turn increase $L_{\mathrm{diff}}$ \cite{Bhaduri18OE,Bhaduri19OL,Hall25OE1km}). Nevertheless, controllably \textit{modulating} $\delta\omega$ enables tuning the axial evolution of one parameter of the STWP without impacting its other characteristics; e.g., tuning $\widetilde{v}$ axially to produce accelerating STWPs \cite{Clerici08OE,Yessenov2020PRLaccel,Li20CP,Li20SR,Li21CP,Hall22OLAccel}. The limits on the tunability of $\widetilde{v}$ elucidated here serve as a blueprint for limits on the achievable axial acceleration. This extends to transverse self-acceleration (bending STWPs \cite{Hall25OL}), and axial spectral encoding whereby the on-axis spectrum controllably shifts with propagation \cite{Motz20arxiv,Hall25OLaxial}. 

\section{Conclusion}

Exploiting recently discovered \textit{non-differentiable AD} opens the way to producing arbitrary group velocities in free space with small NA as long as the pulse bandwidth is sufficiently narrow. Non-differentiable AD requires the minimum NA to realize a wave packet with a given group velocity and bandwidth. In general, dispersive and diffractive effects set the limit on the measured relative group delay with respect to a reference luminal pulse. These results are useful as a benchmark for comparing the utility of any particular model of spatially or spatiotemporally structured optical pulses in the context of modifying their group velocity with fixed resources (pulse bandwidth and NA), which may impact applications in nonlinear \cite{Li22CP}, quantum \cite{Turo26NP}, and topological optics \cite{Yessenov25Meron}.

\medskip
\textbf{Funding} U.S. Office of Naval Research (ONR) N00014-19-1-2192 and N00014-20-1-2789. \\

\textbf{Disclosures} The authors declare no conflicts of interest. \\

\textbf{Data availability} The data that support the findings of this article are not publicly available. The data are available from the author upon reasonable request.

\bibliography{diffraction}

@BOOK{SalehBook07,
  AUTHOR =       {B. E. A. Saleh and M. C. Teich},
  TITLE =        {Fundamentals of Photonics},
  PUBLISHER =    {Wiley},
  YEAR =         {2007},
  address =      {},
}

@ARTICLE{Kondakci16OE,
  AUTHOR =       {H. E. Kondakci and A. F. Abouraddy},
  TITLE =        {Diffraction-free pulsed optical beams via space-time correlations},
  JOURNAL =      {Opt. Express},
  YEAR =         {2016},
  volume =       {24},
  pages =        {28659-28668},
}

@ARTICLE{Saari97PRL,
  AUTHOR =       {P. Saari and K. Reivelt},
  TITLE =        {Evidence of {X}-shaped propagation-invariant localized light waves},
  JOURNAL =      {Phys. Rev. Lett.},
  YEAR =         {1997},
  volume =       {79},
  pages =        {4135-4138},
}

@ARTICLE{Kondakci17NP,
  AUTHOR =       {H. E. Kondakci and A. F. Abouraddy},
  TITLE =        {Diffraction-free space-time light sheets},
  JOURNAL =      {Nat. Photon.},
  YEAR =         {2017},
  volume =       {11},
  pages =        {733-740},
}

@ARTICLE{Kondakci18OE,
  AUTHOR =       {H. E. Kondakci and M. Yessenov and M. Meem and D. Reyes and D. Thul and S. Rostami Fairchild and M. Richardson and R. Menon and A. F. Abouraddy},
  TITLE =        {Synthesizing broadband propagation-invariant space-time wave packets using transmissive phase plates},
  JOURNAL =      {Opt. Express},
  YEAR =         {2018},
  volume =       {26},
  pages =        {13628-13638},
}

@ARTICLE{Yessenov19Optica,
  AUTHOR =       {M. Yessenov and B. Bhaduri and H. E. Kondakci and M. Meem and R. Menon and A. F. Abouraddy},
  TITLE =        {Non-diffracting broadband incoherent space–time fields},
  JOURNAL =      {Optica},
  YEAR =         {2019},
  volume =       {6},
  pages =        {522-607},
}

@ARTICLE{Kondakci18OL,
  AUTHOR =       {H. E. Kondakci and A. F. Abouraddy},
  TITLE =        {Self-healing of space-time light sheets},
  JOURNAL =      {Opt. Lett.},
  YEAR =         {2018},
  volume =       {43},
  pages =        {3830-3833},
}

@ARTICLE{Yessenov19PRA,
  AUTHOR =       {M. Yessenov and B. Bhaduri and H. E. Kondakci and A. F. Abouraddy},
  TITLE =        {Classification of propagation-invariant space-time light-sheets in free space: Theory and experiments},
  JOURNAL =      {Phys. Rev. A},
  YEAR =         {2019},
  volume =       {99},
  pages =        {023856},
}

@ARTICLE{Parker16OE,
  AUTHOR =       {K. J. Parker and M. A. Alonso},
  TITLE =        {The longitudinal iso-phase condition and needle pulses},
  JOURNAL =      {Opt. Express},
  YEAR =         {2016},
  volume =       {24},
  pages =        {28669-28677},
}

@ARTICLE{McGloin05CP,
  AUTHOR =       {D. McGloin and K. Dholakia},
  TITLE =        {Bessel beams: diffraction in a new light},
  JOURNAL =      {Contemp. Phys.},
  YEAR =         {2005},
  volume =       {46},
  pages =        {15-28},
}

@ARTICLE{Mazilu10LPR,
  AUTHOR =       {M. Mazilu and D.J. Stevenson and F. Gunn‐Moore and K. Dholakia},
  TITLE =        {Light beats the spread: “non‐diffracting” beams},
  JOURNAL =      {Laser Photon. Rev.},
  YEAR =         {2010},
  volume =       {4},
  pages =        {529-547},
}

@ARTICLE{SaintMarie17Optica,
  AUTHOR =       {A. Sainte-Marie and O. Gobert and F. Qu{\'e}r{\'e}},
  TITLE =        {Controlling the velocity of ultrashort light pulses in vacuum through spatio-temporal couplings},
  JOURNAL =      {Optica},
  YEAR =         {2017},
  volume =       {4},
  pages =        {1298--1304},
}

@ARTICLE{Bhaduri18OE,
  AUTHOR =       {B. Bhaduri and M. Yessenov and A. F. Abouraddy},
  TITLE =        {Meters-long propagation of diffraction-free space-time light sheets},
  JOURNAL =      {Opt. Express},
  YEAR =         {2018},
  volume =       {26},
  pages =        {20111-20121},
}

@ARTICLE{Bhaduri19OL,
  AUTHOR =       {B. Bhaduri and M. Yessenov and D. Reyes and J. Pena and M. Meem and S. Rostami Fairchild and R. Menon and M. C. Richardson and A. F. Abouraddy},
  TITLE =        {Broadband space-time wave packets propagating 70~m},
  JOURNAL =      {Opt. Lett.},
  YEAR =         {2019},
  volume =       {44},
  pages =        {2073-2076},
}

@ARTICLE{Kondakci19NC,
  AUTHOR =       {H. E. Kondakci and A. F. Abouraddy},
  TITLE =        {Optical space-time wave packets of arbitrary group velocity in free space},
  JOURNAL =      {Nat. Commun.},
  YEAR =         {2019},
  volume =       {10},
  pages =        {929},
}

@ARTICLE{Kondakci19OL,
  AUTHOR =       {H. E. Kondakci and M. A. Alonso and A. F. Abouraddy},
  TITLE =        {Classical entanglement underpins the propagation invariance of space-time wave packets},
  JOURNAL =      {Opt. Lett.},
  YEAR =         {2019},
  volume =       {44},
  pages =        {2645-2648},
}

@ARTICLE{Giovannini15Science,
  AUTHOR =       {D. Giovannini and J. Romero and V. Poto{\v c} and G. Ferenczi and F. Speirits and S. M. Barnett and D. Faccio and M. J. Padgett},
  TITLE =        {Spatially structured photons that travel in free space slower than the speed of light},
  JOURNAL =      {Science},
  YEAR =         {2015},
  volume =       {347},
  pages =        {857-860},
}

@ARTICLE{Bouchard16Optica,
  AUTHOR =       {F. Bouchard and J. Harris and H. Mand and R. W. Boyd and E. Karimi},
  TITLE =        {Observation of subluminal twisted light in vacuum},
  JOURNAL =      {Optica},
  YEAR =         {2016},
  volume =       {3},
  pages =        {351-354},
}

@ARTICLE{Froula18NP,
  AUTHOR =       {D. H. Froula and D. Turnbull and A. S. Davies and T. J. Kessler and D. Haberberger and J. P. Palastro and S.-W. Bahk and I. A. Begishev and R. Boni and S. Bucht and J. Katz and J. L. Shaw},
  TITLE =        {Spatiotemporal control of laser intensity},
  JOURNAL =      {Nat. Photon.},
  YEAR =         {2018},
  volume =       {12},
  pages =        {262-265},
}

@ARTICLE{Zapata06OL,
  AUTHOR =       {C. J. Zapata-Rodr{\'i}guez and M. A. Porras},
  TITLE =        {X-wave bullets with negative group velocity in vacuum},
  JOURNAL =      {Opt. Lett.},
  YEAR =         {2006},
  volume =       {31},
  pages =        {3532-3534},
}

@ARTICLE{Clerici08OE,
  AUTHOR =       {M. Clerici and D. Faccio and A. Lotti and E. Rubino and O. Jedrkiewicz and J. Biegert and P. Di Trapani},
  TITLE =        {Finite-energy, accelerating {B}essel pulses},
  JOURNAL =      {Opt. Express},
  YEAR =         {2008},
  volume =       {16},
  pages =        {19807-19811},
}

@ARTICLE{Porras03PRE2,
  AUTHOR =       {M. A. Porras and G. Valiulis and P. {Di T}rapani},
  TITLE =        {Unified description of {B}essel {X} waves with cone dispersion and tilted pulses},
  JOURNAL =      {Phys. Rev. E},
  YEAR =         {2003},
  volume =       {68},
  pages =        {016613},
}

@ARTICLE{Yessenov19OE,
  AUTHOR =       {M. Yessenov and B. Bhaduri and L. Mach and D. Mardani and H. E. Kondakci and M. A. Alonso and G. A. Atia and A. F. Abouraddy},
  TITLE =        {What is the maximum differential group delay achievable by a space-time wave packet in free space?},
  JOURNAL =      {Opt. Express},
  YEAR =         {2019},
  volume =       {27},
  pages =        {12443-12457},
}

@ARTICLE{Yessenov19OL,
  AUTHOR =       {M. Yessenov and A. F. Abouraddy},
  TITLE =        {Changing the speed of coherence in free space},
  JOURNAL =      {Opt. Lett.},
  YEAR =         {2019},
  volume =       {44},
  pages =        {5125-5128},
}

@article{Schepler20ACSPhot,
  title={Space--time surface plasmon polaritons: A new propagation-invariant surface wave packet},
  author={Schepler, K. L and Yessenov, M. and Zhiyenbayev, Y. and Abouraddy, A. F},
  journal={ACS Photon.},
  volume={7},
  pages={2966--2977},
  year={2020},
}

@ARTICLE{Bhaduri20NP,
  AUTHOR =       {B. Bhaduri and M. Yessenov and A. F. Abouraddy},
  TITLE =        {Anomalous refraction of optical space-time wave packets},
  JOURNAL =      {Nat. Photon.},
  YEAR =         {2020},
  volume =       {14},
  pages =        {416-421},
}

@ARTICLE{Hall21OL,
  AUTHOR =       {L. A. Hall and M. Yessenov and A. F. Abouraddy},
 title={Space--time wave packets violate the universal relationship between angular dispersion and pulse-front tilt},
  journal={Opt. Lett.},
  volume={46},
  number={7},
  pages={1672--1675},
  year={2021},
  publisher={Optical Society of America}
}

@article{Yessenov21ACSPhot,
author = {Yessenov, M. and Hall, L. A. and Abouraddy, A. F.},
title = {Engineering the Optical Vacuum: Arbitrary Magnitude, Sign, and Order of Dispersion in Free Space Using Space–Time Wave Packets},
journal = {ACS Photonics},
volume = {8},
number = {8},
pages = {2274-2284},
year = {2021},
doi = {10.1021/acsphotonics.1c00275},
URL = {         https://doi.org/10.1021/acsphotonics.1c00275},
}

@ARTICLE{Motz20arxiv,
  AUTHOR =       {A. M. {Allende Motz} and M. Yessenov and A. F. Abouraddy},
  TITLE =        {Axial spectral encoding of space-time wave packets},
  JOURNAL =      {Phys. Rev. Appl.},
  YEAR =         {2021},
  volume =       {15},
  pages =        {024067},
}

@ARTICLE{Malaguti08OL,
  AUTHOR =       {S. Malaguti and G. Bellanca and S. Trillo},
  TITLE =        {Two-dimensional envelope localized waves in the anomalous dispersion regime},
  JOURNAL =      {Opt. Lett.},
  YEAR =         {2008},
  volume =       {33},
  pages =        {1117-1119},
}

@ARTICLE{Malaguti09PRA,
  AUTHOR =       {S. Malaguti and S. Trillo},
  TITLE =        {Envelope localized waves of the conical type in linear normally dispersive media},
  JOURNAL =      {Phys. Rev. A},
  YEAR =         {2009},
  volume =       {79},
  pages =        {063803},
}

@article{Yessenov21JOSAA3,
author = {M. Yessenov and A. M. {Allende Motz} and B. Bhaduri and A. F. Abouraddy},
journal = {J. Opt. Soc. Am. A},
number = {10},
pages = {1462--1470},
publisher = {OSA},
title = {Refraction of space-time wave packets: {III}. experiments at oblique incidence},
volume = {38},
month = {Oct},
year = {2021},
url = {http://josaa.osa.org/abstract.cfm?URI=josaa-38-10-1462},
doi = {10.1364/JOSAA.430109},
}

@Article{Motz21OL,
  author   = {Allende Motz, A. M. and Yessenov, M. and Abouraddy, A. F.},
  journal  = {Opt. Lett.},
  title    = {Isochronous space–time wave packets},
  year     = {2021},
  pages    = {2260-2263},
  volume   = {46},
}

@Article{Hall21APLSTTalbot,
  author  = {L. A. Hall and M. Yessenov and S. A. Ponomarenko and A. F. Abouraddy},
  journal = {APL Photon.},
  title   = {The space-time {T}albot effect},
  year    = {2021},
  pages   = {056105},
  volume  = {6},
}

@Article{Yessenov2020PRLaccel,
  author    = {Yessenov, M. and Abouraddy, A. F.},
  journal   = {Phys. Rev. Lett.},
  title     = {Accelerating and decelerating space-time optical wave packets in free space},
  year      = {2020},
  pages     = {233901},
  volume    = {125},
}

@article{Yessenov22AOP,
    title={Space-time wave packets},
    author={Yessenov, M. and Hall, L. A. and Schepler, K. L. and Abouraddy, A. F.},
    journal = {Adv. Opt. Photon.},
    pages = {455-570},
    volume = {14},
    year = {2022},
}

@article{Yessenov22NC,
  title={Space-time wave packets localized in all dimensions},
  author={M. Yessenov and J. Free and Z. Chen and E. G. Johnson and M. P. J. Lavery and M. A. Alonso and A. F. Abouraddy},
  journal={Nat. Commun.},
  volume={13},
  pages={4573},
  year={2022},
}

@ARTICLE{Hall23NatPhys,
  AUTHOR =       {L. A. Hall and A. F. Abouraddy},
  TITLE =        {Observation of optical {de B}roglie-{M}ackinnon wave packets},
  JOURNAL =      {Nat. Phys.},
  YEAR =         {2023},
  volume =       {19},
  pages =        {435-444},
}

@ARTICLE{Yessenov22OL,
  AUTHOR =       {M. Yessenov and Z. Chen and M. P. J. Lavery and A. F. Abouraddy},
  TITLE =        {Vector space-time wave packets},
  JOURNAL =      {Opt. Lett.},
  YEAR =         {2022},
  volume =       {47},
  pages =        {4131-4134},
}

@ARTICLE{Diouf22SA,
  AUTHOR =       {M. Diouf and Z. Lin and M. Harling and  K. C. Toussaint},
  TITLE =        {Demonstration of speckle resistance using space–time light sheets},
  JOURNAL =      {Sci. Rep.},
  YEAR =         {2022},
  volume =       {12},
  pages =        {14064},
}

@ARTICLE{Ramsey23PRA,
  AUTHOR =       {D. Ramsey and A. Di Piazza and M. Formanek and P. Franke and D. H. Froula and B. Malaca and W. B. Mori and J. R. Pierce and T. T. Simpson and J. Vieira and M. Vranic and K. Weichman and J. P. Palastro},
  TITLE =        {Exact solutions for the electromagnetic fields of a flying focus},
  JOURNAL =      {Phys. Rev. A},
  YEAR =         {2023},
  volume =       {107},
  pages =        {013513},
}

@ARTICLE{Hall22OEConsequences,
  AUTHOR =       {L. A. Hall and A. F. Abouraddy},
  TITLE =        {Consequences of non-differentiable angular dispersion in optics: tilted pulse fronts versus space-time wave packets},
  JOURNAL =      {Opt. Express},
  YEAR =         {2022},
  volume =       {30},
  pages =        {4817-4832},
}

@ARTICLE{Hall22JOSAA,
  AUTHOR =       {L. A. Hall and A. F. Abouraddy},
  TITLE =        {Non-differentiable angular dispersion as an optical resource},
  JOURNAL =      {J. Opt. Soc. Am. A},
  YEAR =         {2022},
  volume =       {39},
  pages =        {2016-2025},
}

@ARTICLE{Almeida25OL,
  AUTHOR =       {R. R. Almeida and D. Ramsey and A. F. Abouraddy and J. P. Palastro and J. Vieira},
  TITLE =        {Universal structure of propagation-invariant optical pulses},
  JOURNAL =      {Opt. Lett.},
  YEAR =         {2025},
  volume =       {50},
  pages =        {3393-3396},
}

@ARTICLE{Hall24ACSP,
  AUTHOR =       {L. A. Hall and A. F. Abouraddy},
  TITLE =        {Abrupt {X}-to-{O}-wave structural field transition in presence of anomalous dispersion},
  JOURNAL =      {APL Photon.},
  YEAR =         {2024},
  volume =       {9},
  pages =        {126106},
}

@ARTICLE{Hall23LPR,
  AUTHOR =       {L. A. Hall and A. F. Abouraddy},
  TITLE =        {Canceling and inverting normal and anomalous group-velocity dispersion using space-time wave packets},
  JOURNAL =      {Laser Photon. Rev.},
  YEAR =         {2023},
  volume =       {17},
  pages =        {2200119},
}

@ARTICLE{Hall25PRA,
  AUTHOR =       {L. A. Hall and A. F. Abouraddy},
  TITLE =        {Spectral reorganization of space-time wave packets in presence of normal group-velocity dispersion},
  JOURNAL =      {Phys. Rev. A},
  YEAR =         {2025},
  volume =       {111},
  pages =        {063515},
}

@ARTICLE{Hall21OLNormalGVD,
  AUTHOR =       {L. A. Hall and A. F. Abouraddy},
  TITLE =        {Realizing normal group-velocity dispersion in free space via angular dispersion},
  JOURNAL =      {Opt. Lett.},
  YEAR =         {2021},
  volume =       {46},
  pages =        {5421-5424},
}

@ARTICLE{Hall21PRAVwave,
  AUTHOR =       {L. A. Hall and A. F. Abouraddy},
  TITLE =        {Free-space group-velocity dispersion induced in space-time wave packets by {V}-shaped spectra},
  JOURNAL =      {Phys. Rev. A},
  YEAR =         {2021},
  volume =       {104},
  pages =        {013505},
}

@ARTICLE{Yessenov25Meron,
  AUTHOR =       {M. Yessenov and A. H. Dorrah and C. Guo and L. A. Hall and J.-S. Park and J. Free and E. G. Johnson and F. Capasso and S. Fan and A. F. Abouraddy},
  TITLE =        {Ultrafast space-time optical merons in momentum-energy space},
  JOURNAL =      {Nat. Commun.},
  YEAR =         {2025},
  volume =       {16},
  pages =        {8592 },
}

@ARTICLE{Hall24JOSAA,
  AUTHOR =       {L. A. Hall and A. F. Abouraddy},
  TITLE =        {Universal angular-dispersion synthesizer},
  JOURNAL =      {J. Opt. Soc. Am. A},
  YEAR =         {2024},
  volume =       {41},
  pages =        {83-94},
}

@article{Yessenov20PRL,
  author =      {M. Yessenov and L. A. Hall and S. A. Ponomarenko and A. F. Abouraddy},
  title =       {Veiled {T}albot effect},
  journal =     {Phys. Rev. Lett.},
  volume =      {125},
  pages =       {243901},
  year =        {2020},
}

@ARTICLE{Hall21OLTalbot,
  AUTHOR =       {L. A. Hall and S. Ponomarenko and A. F. Abouraddy},
  TITLE =        {Temporal {T}albot effect in free space},
  JOURNAL =      {2021},
  YEAR =         {Opt. Lett.},
  volume =       {46},
  pages =        {3107-3110},
}

@ARTICLE{Li20CP,
  AUTHOR =       {Z. Li and J. Kawanaka},
  TITLE =        {Optical wave-packet with nearly-programmable group velocities},
  JOURNAL =      {Commun. Phys.},
  YEAR =         {2020},
  volume =       {3},
  pages =        {211},
}

@ARTICLE{Li21CP,
  AUTHOR =       {Z. Li and Y. Gu and J. Kawanaka},
  TITLE =        {Reciprocating propagation of laser pulse intensity in free space},
  JOURNAL =      {Commun. Phys.},
  YEAR =         {2021},
  volume =       {4},
  pages =        {87},
}

@ARTICLE{Li20SR,
  AUTHOR =       {Z. Li and J. Kawanaka},
  TITLE =        {Velocity and acceleration freely tunable straight-line propagation light bullet},
  JOURNAL =      {Sci. Rep.},
  YEAR =         {2020},
  volume =       {10},
  pages =        {11481},
}

@ARTICLE{Hall22OLAccel,
  AUTHOR =       {L. A. Hall and M. Yessenov and A. F. Abouraddy},
  TITLE =        {Arbitrarily accelerating space-time wave packets},
  JOURNAL =      {Opt. Lett.},
  YEAR =         {2022},
  volume =       {47},
  pages =        {694-697},
}

@ARTICLE{Hall25OE1km,
  AUTHOR =       {L. A. Hall and M. A. Romer and B. L. Turo and T. M. Hayward and R. Menon and A. F. Abouraddy},
  TITLE =        {Space-time wave packets propagating a kilometer in air},
  JOURNAL =      {arXiv:2209.03309},
  YEAR =         {2022},
  volume =       {},
  pages =        {},
}

@ARTICLE{Pigeon24OE,
  AUTHOR =       {J. J. Pigeon and P. Franke and M. {Lim Pac C}hong and J. Katz and R. Boni and C. Dorrer and J. P. Palastro and D. H. Froula},
  TITLE =        {Ultrabroadband flying-focus using an axiparabola-echelon pair},
  JOURNAL =      {Opt. Express},
  YEAR =         {2024},
  volume =       {32},
  pages =        {576-585},
}

@ARTICLE{Romer25JOpt,
  AUTHOR =       {M. A. Romer and L. A. Hall and A. F. Abouraddy},
  TITLE =        {Synthesis and characterization of space-time light sheets: a tutorial},
  JOURNAL =      {J. Opt.},
  volume =       {27},
  pages =        {013501},
}

@ARTICLE{Sambles15Science,
  AUTHOR =       {J. R. Sambles},
  TITLE =        {Structured photons take it slow},
  JOURNAL =      {Science},
  YEAR =         {2015},
  volume =       {347},
  pages =        {828},
}

@ARTICLE{Hall25APLP,
  AUTHOR =       {L. A. Hall and M. Yessenov and K. L. Schepler and A. F. Abouraddy},
  TITLE =        {Universality and Non-differentiability: {A} new perspective on angular dispersion in optics},
  JOURNAL =      {APL Photon.},
  YEAR =         {2025},
  volume =       {10},
  pages =        {121101},
}

@article{Ichiji25NC,
  author    = {Ichiji, Naoki and Kikuchi, Hibiki and Yessenov, Murat and Schepler, Kenneth L. and Abouraddy, Ayman F. and Kubo, Atsushi},
  title     = {Observation of space-time surface plasmon polaritons},
  journal   = {Nature Comm.},
  volume    = {16},
  pages     = {10697},
  year      = {2025},
}

@article{Hall25OL,
  author    = { Layton A. Hall and Ayman F. Abouraddy},
  title     = {Bending space-time wave packets},
  journal   = {Opt. Lett.},
  volume    = {50},
  number    = {18},
  pages     = {5829},
  year      = {2025},
}

@article{Hall25OLaxial,
  author    = {L. A. Hall and M. Yessenov and M. A. Romer and A. F. Abouraddy},
  title     = {Long-distance axial spectral encoding using space-time wave packets},
  journal   = {Opt. Lett.},
  volume    = {50},
  number    = {15},
  pages     = {4698},
  year      = {2025},
}

@BOOK{Kurgin08Book,
  Editor =       {J. B. Khurgin and R. S. Tucker},
  TITLE =        {Slow Light: {S}cience and Applications},
  YEAR =         {2008},
  publisher =    {CRC Press},
  address =      {Boca Raton, FL},
}

@ARTICLE{Martinez84JOSAA,
  AUTHOR =       {O. E. Martinez and J. P. Gordon and R. L. Fork},
  TITLE =        {Negative group-velocity dispersion using refraction},
  JOURNAL =      {J. Opt. Soc. Am. A},
  YEAR =         {1984},
  volume =       {1},
  pages =        {1003-1006},
}

@ARTICLE{Yessenov25JOSAA,
  AUTHOR =       {M. Yessenov and A. F. Abouraddy},
  TITLE =        {Optical spatiotemporal {F}ourier synthesis: tutorial},
  JOURNAL =      {J. Opt. Soc. Am. A},
  YEAR =         {2025},
  volume =       {42},
  pages =        {1295-1315},
}

@ARTICLE{Diouf23Optica,
  AUTHOR =       {M. Diouf and Z. Lin and M. Harling and K. Krishna and K. C. Toussaint},
  TITLE =        {Interferometric phase stability from Gaussian and space--time light sheets},
  JOURNAL =      {Optica},
  YEAR =         {2023},
  volume =       {10},
  pages =        {1161-1164},
}

@article{Ichiji24JOSAA,
    author = {N. Ichiji and H. Kikuchi and M. Yessenov and K. L. Schepler and A. F. Abouraddy and A. Kubo},
    title = {Exciting space-time surface plasmon polaritons by irradiating a nanoslit structure},
    journal = {J. Opt. Soc. Am. A},
    year = {2024},
    volume = {41},
    pages = {396-405},
}

@article{Ichiji23PRA,
    author = {N. Ichiji and D. Oue and M. Yessenov and K. L. Schepler and A. F. Abouraddy and A. Kubo},
    title = {Transverse spin angular momentum of a space-time surface plasmon polariton wave packet},
    journal = {Phys. Rev. A},
    year = {2023},
    volume = {107},
    pages = {063517},
}

@article{Ichiji23ACSP,
    author = {N. Ichiji and H. Kikuchi and M. Yessenov and K. L. Schepler and A. F. Abouraddy and A. Kubo},
    title = {Observation of ultrabroadband striped space-time surface plasmon polaritons},
    journal = {ACS Photon.},
    year = {2023},
    volume = {10},
    pages = {374-382},
}

@ARTICLE{Zhang25LSM,
  AUTHOR =       {J. Zhang and H. Luo and D. M. Mitchell and S. Luckhart and M. Khajavikhan and A. F. Abouraddy and D. N. Christodoulides and A. E. Vasdekis},
  TITLE =        {Space-time light-sheet microscopy},
  JOURNAL =      {bioRxiv 2026.04.10.717581},
  YEAR =         {2026},
  volume =       {},
  pages =        {},
}

@ARTICLE{Hall26SelectiveAvoidance,
  AUTHOR =       {L. A. Hall and M. Yessenov and I. Lebron and A. F. Abouraddy},
  TITLE =        {Selective avoidance of multiple line-of-sight obstacles at 130~m using locally bending space-time wave packets},
  JOURNAL =      {arXiv:2607.12311},
  YEAR =         {2026},
  volume =       {},
  pages =        {},
}

@ARTICLE{Hall26SubRayleigh,
  AUTHOR =       {L. A. Hall and B. L. Turo and A. F. Abouraddy},
  TITLE =        {Avoiding a line-of-sight obstacle via deep sub-{R}ayleigh shadow-projection using space-time wave packets},
  JOURNAL =      {arXiv:2607.02752},
  YEAR =         {2026},
  volume =       {},
  pages =        {},
}

@ARTICLE{Turo26NP,
  AUTHOR =       {B. L. Turo and B. E. A. Saleh and A. F. Abouraddy},
  TITLE =        {Local and remote synthesis of single-photon space-time wave packets},
  JOURNAL =      {Nat. Photon.},
  YEAR =         {2026},
  volume =       {20},
  pages =        {539-548},
}

@article{Li22CP,
    author = {Z. Li and Y. Leng and R. Li},
    title = {Strong double space-time wave packets using optical parametric amplification},
    journal = {Commun. Phys.},
    year = {2022},
    volume = {5},
    pages = {295},
}

@ARTICLE{Hall26JOSAA1,
  AUTHOR =       {L. A. Hall and A. F. Abouraddy},
  TITLE =        {Limits on the free-space group velocity of optical wave packets incorporating angular dispersion. {Part~I}, conventional angular dispersion: tutorial},
  JOURNAL =      {accompanying paper},
  YEAR =         {2026},
  volume =       {},
  pages =        {},
}

\end{document}